\documentclass[aps,amsmath,amssymb,twocolumn]{revtex4}
\usepackage{amsmath,amssymb}
\usepackage{graphicx}
\usepackage{dcolumn}
\usepackage{bm}
\usepackage{tabularx}
\usepackage{booktabs}

  \def\g{{\gamma}} 
  
 \def\q{{\bar q}} 
  \def\u{{\bar u}}
  \def\d{{\bar d}}

   \def\Ci{C^{-1}}

  \def\gt_#1{(-C\g_{#1}\Ci)}
  \def\bra{\langle}
  \def\ket{\rangle}
  \def\qcon{\bra\q q\ket}
  \def\uc{\bra\u u\ket}
  \def\dc{\bra\d d\ket}
  
  \def\gc{\left\bra \frac{\alpha_s}{\pi}G^2\right\ket}
  \def\mixq{g\left\bra\q\sigma\cdot G q\right\ket}
  \def\mixu{g\left\bra\u\sigma\cdot G u\right\ket}
  \def\mixd{g\left\bra\d\sigma\cdot G d\right\ket}

\def\ben{\begin{equation}}
\def\een{\end{equation}}
\def\bey{\begin{eqnarray}}
\def\eey{\end{eqnarray}}

  \def\bra{\langle}
  \def\ket{\rangle}
  \def\psla{p{\raise1pt\hbox{$\!\!/$}}}
  \def\qsla{q{\raise1pt\hbox{$\!\!\!/$}}}   
  \def\ksla{k{\raise1pt\hbox{$\!\!\!/$}}}
  \def\vsla{v{\raise1pt\hbox{$\!\!\!/$}}}
  \def\Dsla{D{\raise1pt\hbox{$\!\!\!\!/$}}}

    \def\vecv{\textnormal{\mathversion{bold}$v$}}
     
     \def\vec0{\textnormal{\mathversion{bold}$0$}}

  \def\eps{\epsilon}
  \def\psla{p{\raise1pt\hbox{$\!\!/$}}}
  \def\qsla{q{\raise1pt\hbox{$\!\!\!/$}}}
  \def\knd(#1,#2){\delta_{#1#2}}
    
  \def\vecv{\textnormal{\mathversion{bold}$v$}}

   \def\Tr{{\rm Tr}}

\def\aplt{\kern0.3333em \raise 0.2ex \hbox{$<$}%
\kern-0.8em \lower0.8ex \hbox{$\sim$}%
\kern0.3333em}
\def\aplg{\kern0.3333em \raise 0.2ex \hbox{$>$}%
\kern-0.8em \lower0.8ex \hbox{$\sim$}%
\kern0.3333em}
\def\ben{\begin{equation}}
\def\een{\end{equation}}
\def\bey{\begin{eqnarray}}
\def\eey{\end{eqnarray}}
\def\ba{\begin{array}}
\def\ea{\end{array}}
\def\benmrt{\begin{enumerate}}
\def\eenmrt{\end{enumerate}}

\def\psla{p{\raise1pt\hbox{$\!\!\!/$}}}
\def\dsla{\partial{\raise1pt\hbox{$\!\!\!/$}}}
\def\Dsla{D{\raise1pt\hbox{$\!\!\!\!/$}}}
\def\xsla{x{\raise1pt\hbox{$\!\!\!/$}}}

\def\jmu5{j_{\mu 5}^{(i)}(0)}
\def\jnu5{j_{\nu 5}^{(i)}(0)}

\def\qq0v{\langle0\!\mid\!{\bar q}q\!\mid\! 0\rangle}

\def\qc0f{\langle0\!\mid\!{\bar q}q\!\mid\!0\rangle_{F}}

\def\qsq0f{\langle0\!\mid\!{\bar q}\sigma_{\mu\nu}q\!\mid\!0\rangle_{F}}

\def\qgdq0f{\langle0\!\mid\!{\bar q}{\cal 
S}\gamma_{\mu}D_{\nu}q\!\mid\!0\rangle_{F}}

\def\qddq0f{\langle0\!\mid\!{\bar q}{\cal
S}D_{\mu}D_{\nu}q\!\mid\!0\rangle_{F}}

\def\3mmtm{|{\bf q}|^2}

\def\eq#1{Eq.(\ref{#1})}
\def\eqs#1#2{Eqs.(\ref{#1}) and (\ref{#2})}

\def\meq#1#2{Eqs.(\ref{#1})$\sim$(\ref{#2})}

\def\Ref#1{[\ref{#1}]}
\def\Refs#1#2{[\ref{#1},\ref{#2}]}

\def\p0{p_0}

\def\gam3{\mbox{\boldmath{$\gamma$}}}
\def\e0{E_{0}(s_{0},s)}
\def\e1{E_{1}(s_{0},s)}
\def\e2{E_{2}(s_{0},s)}

\def\aplt{\kern0.3333em \raise 0.2ex \hbox{$<$}%
\kern-0.8em \lower0.8ex \hbox{$\sim$}%
\kern0.3333em}
\def\aplg{\kern0.3333em \raise 0.2ex \hbox{$>$}%
\kern-0.8em \lower0.8ex \hbox{$\sim$}%
\kern0.3333em}

\def\ln{{\rm ln}}

\begin{document}
\preprint{}
\title{QCD sum rules for positive and negative parity heavy baryons at next-to-leading order in $\alpha_s$-expansion}

\author{Tetsuo NISHIKAWA}
\email{nishikawa@sbctmu.ac.jp} 
\affiliation{%
Faculty of Health Science, SBC Tokyo Medical University, Urayasu, Chiba 279-8567, Japan
}
\author{Yoshihiko KONDO}
\email{kondo@kokugakuin.ac.jp} 
\affiliation{%
Faculty of Human development, Kokugakuin University, Yokohama 225-0003, Japan
}
\author{Yoshiko Kanada-En'yo}
\affiliation{%
Department of Physics, Kyoto University, Kyoto 606-8502, Japan 
}

\date{\today}

\begin{abstract}
QCD sum rules for positive and negative parity heavy baryons in the heavy quark limit are formulated.
We apply the method to $\Lambda$ and $\Sigma$ channels.
We include the next-to-leading order corrections in $\alpha_s$-expansion to dimension 0 and 3 terms in the operator product expansion.
The corrections lead to the considerable reduction of the predicted masses and significantly improves the stability with respect to the Borel parameter, especially for negative parity states.
It is also found that in the heavy quark limit
chiral odd condensates do not contribute
to the negative parity states.
\end{abstract}
\pacs{
}
\maketitle

\section{Introduction}
\label{sec1}
In the past decades, a remarkable experimental progress has been made in the field of heavy baryon physics.
In fact, many excited states of singly charmed baryon have been observed in 2000's.
Excited singly bottom baryons have also been discovered in recent years one after another,
although only two states were known until 2012 \cite{zhureview1,zhureview2}. 
Those data have been investigated comprehensively from various theoretical perspective so far (see \cite{decaySR} and references therein).
To study negative parity heavy baryons is especially important in the sense that they can be key subjects to clarify the mechanism of excitation in baryon systems. 

QCD sum rule is one of the useful nonperturbative method based on QCD, which can connect the nontrivial vacuum condensates with the hadron properties in a model independent way. 
So far, by many authors QCD sum rule has been used to study not only ground (positive parity) \cite{shuryak,grozinyakovlev,BCDN,dai,Groote1,Groote2} but also excited (negative parity) heavy baryons \cite{leeliusong,huangzhangzhu,wanghuang,liuchenliuhosakazhu,chenchenmaohosakaliuzhu,maochenchenhosakaliuzhu,decaySR}.
The authors in Ref.\cite{leeliusong,huangzhangzhu,wanghuang,liuchenliuhosakazhu,chenchenmaohosakaliuzhu,maochenchenhosakaliuzhu,decaySR} systematically constructed the $p$-wave heavy baryon interpolating fields with a covariant derivative and applied them to study the mass spectrum and decay properties of $p$-wave heavy baryons \cite{decaySR}.

In this paper we employ a completely different approach that was originally proposed in Ref.\cite{JKO}.
We use the interpolating field of positive intrinsic parity without covariant derivatives.
Noting the fact that the interpolating field of positive intrinsic parity couples not only to positive parity states but also to negative parity ones, we \lq\lq project" the correlation function of the interpolating fields onto each parity \cite{JKO}, which enables us to construct the sum rules for respective parity states. Nucleons and hyperons were investigated within this method and the origin of the mass splitting between positive and negative parity was discussed in connection with chiral condensates \cite{JKO,JO,KMN1,KMN2}. Up to now, there exist no work applying properly this method to studying heavy baryons.

We consider heavy baryons containing one heavy quark and construct the QCD sum rules in the framework of the heavy quark effective theory (HQET), since the physics of hadrons containing one infinitely heavy quark is well described with HQET
and the analysis is greatly simplified due to the heavy quark symmetry.


We take into account the next-to-leading order corrections in $\alpha_s$ to the terms in the operator product expansion.
It is known that in the systems containing heavy quarks $\alpha_s$-correction gives significantly large contribution, which amounts to 100\% or more to the leading order contribution \cite{bagan,Groote1,Groote2,Bmeson3body}.

The paper is organized as follows.
In the second section, we formulate the QCD sum rules for positive and negative parity heavy baryons. The method is applied to $\Lambda$ and $\Sigma$ heavy baryons and demonstrate the derivation of the sum rules at leading order in $\alpha_s$, in the third section.
In the fourth section, we calculate the next-to-leading order correction in $\alpha_s$
and present the sum rules including the corrections.
The sum rules derived in the third and fourth sections are analyzed numerically in the fifth section. The sixth section is devoted to summary and discussion.

\section{QCD sum rules for positive and negative parity heavy baryons in the HQET}
\label{sec2}
In this section, we set up the QCD sum rule for positive and negative parity heavy baryons in the HQET. 
We consider the following correlation function in the HQET:
\bey
\Pi_B(\omega)&=&i\int d^4x e^{i\omega v\cdot x}\bra 0|T\left[\eta_B(x)\bar\eta_B(0)\right]|0\ket,
\label{hbcorfun}
\eey
where $\eta_B$ is the interpolating field of heavy baryon $B$ and $v$ is the four velocity of the heavy baryon.
In the general procedures in the QCD sum rule approach, the correlation functions are evaluated by the operator product expansion (OPE) in unphysical region $\omega\rightarrow -\infty$ on one hand and expressed in terms of the properties of physical states (masses, coupling constants and so on) on the other hand; we relate the two descriptions exploiting the dispersion relations, which yields the QCD sum rule.

Let us first consider how \eq{hbcorfun} is expressed in terms of physical states.
The interpolating field $\eta_B$ couples not only to positive parity states but also to negative party ones \cite{CDKS,JKO}, in the way that
\bey
&&\bra 0|\eta(0)|B_{j(+)}(v,\alpha)\ket=\lambda_{j(+)} u(v,\alpha),
\label{couplingtopositive}\\
&&\bra 0|\eta(0)|B_{j(-)}(v,\alpha)\ket=\lambda_{j(-)} \gamma_5 u(v,\alpha),
\label{couplingtonegative}
\eey
where $|B_{j(\pm)}(v,\alpha)\ket$ is the $j$-th positive/negative parity resonance state with velocity $v$ and spin $\alpha$ and $u(v,\alpha)$ is the Dirac spinor for the baryon at heavy quark limit. 
Hence, inserting a complete set of physical states between the two interpolating fields in \eq{hbcorfun} and neglecting the widths of the resonance states, \eq{hbcorfun} in the baryon rest frame $\vecv=\vec0$ can be expressed as follows, 
\bey
\Pi(\omega)&=&\sum_{j}\left[
\frac{-|\lambda_{j(+)}|^2}{\omega-\bar{\Lambda}_{j(+)}+i\epsilon}P_+
+\frac{-|\lambda_{j(+)}|^2}{\omega-\bar{\Lambda}_{j(-)}+i\epsilon}P_-\right],
\nonumber\\
&&
\label{corfunphysical}
\eey
where $P_\pm\equiv\frac{1}{2}(\gamma_0\pm1)$ are the parity projection operators and $\bar\Lambda_{j(\pm)}\equiv M_{j(\pm)}-m_Q$ with $M_{j(\pm)}$ being the masses of $j$-th positive/negative parity states of the heavy baryon and $m_Q$ the heavy quark mass.
It should be noticed that \eq{corfunphysical} does not have poles at negative $\omega$ because of the absence of anti-heavy baryons in the heavy quark limit. In the correlation function of nucleons or hyperons, the term corresponding to the first (second) one in \eq{corfunphysical} has poles of anti-particles of negative (positive) parity at negative $\omega$. As a result, even after performing \lq\lq parity projection" (see below), the contributions of positive and negative parity states are not separated from each other \cite{KMN1,KMN2}.

We apply \lq\lq parity projection" onto the correlation function, namely, consider
\bey
\Pi_{B(\pm)}(\omega)=\frac{1}{4}\Tr\left[P_{\pm}\Pi(\omega)\right].
\label{projectedcorfun}
\eey
From \eq{corfunphysical} we see $\Pi_{B(+)}(\omega)$ contain only the contribution of the positive parity states and $\Pi_{B(-)}(\omega)$ only the negative parity.
Calculating \eq{projectedcorfun} using the OPE and matching the results with the corresponding parity components in \eq{corfunphysical}, we obtain the sum rules.
The matching can be done via the dispersion relations for $\Pi_{B(\pm)}$ and utilizing Borel transformation.
$\Pi_{B(\pm)}$ obeys the the dispersion relations,
\ben
\Pi_{B(\pm)}(\omega)=\int^\infty_{0}d\omega'\frac{\rho_{B(\pm)}(\omega')}{\omega'-\omega-i\epsilon},
\label{disprela}
\een
(the subtraction terms, which are polynomial in $\omega$, are not written explicitly here).
In \eq{disprela}, $\rho_{B(\pm)}(\omega)$ is the spectral function defined by
\bey
\rho_{B(\pm)}(\omega)=\frac{1}{\pi}{\rm Im}\Pi_{B(\pm)}(\omega),
\eey
We apply the Borel transformation operator, defined by
\bey
\hat{B}\equiv\lim_{\stackrel{n\rightarrow\infty,-\omega\rightarrow\infty}{M=-\omega/n\ {\rm fixed}}}
\frac{\omega^n}{\Gamma(n)}\left(-\frac{d}{d\omega}\right)^n,
\label{Borel}
\eey
on both sides of the dispersion relation, \eq{disprela}.
This transformation introduces an exponential weight in the integral as
\bey
\hat{B}\Pi_{B(\pm)}(\omega)=\frac{1}{M}\int^\infty_{0}d\omega'
e^{-\omega'/M}
\rho_{B(\pm)}(\omega'),
\label{disprelaBorel}
\eey
and eliminate the subtraction terms. 
In the LHS of \eq{disprelaBorel}, where $\Pi_{B(\pm)}(\omega)$ are calculated by OPE, 
the convergence of the series in the OPE is improved since the higher dimensional terms in the OPE are  suppressed factorially ($\sim 1/n!$).
Simultaneously, in the RHS, the contributions of higher resonances and continuum are suppressed exponentially compared with that of the lowest-lying state.
It is therefore allowed to approximate excited-state contributions 
to the RHS of \eq{disprelaBorel} by the imaginary part of the OPE result which starts from the \lq\lq continuum threshold" 
$\omega_{\rm th}$; namely, we use
\bey
\rho_{B(\pm)}(\omega)
&=&|\lambda_{B(\pm)}|^2\delta(\omega-\bar\Lambda_{B(\pm)})\cr
&&+\frac{1}{\pi}{\rm Im}\Pi_{B(\pm)}^{\rm OPE}(\omega)\theta(\omega-\omega_{\rm th}),
\label{spectralansatz}
\eey
where $\Pi_{B(\pm)}^{\rm OPE}(\omega)$ is the correlation function calculated by OPE.
Substituting \eq{spectralansatz} into the RHS of \eq{disprelaBorel}, we obtain,
\bey
|\lambda_{B(\pm)}|^2 e^{-\bar\Lambda_{B(\pm)}/M} = 
\int_{0}^{\omega_{\rm th}} d\omega e^{-\omega/M}\frac{1}{\pi}{\rm Im}\Pi_{B(\pm)}^{\rm OPE}(\omega),
\cr&&
\label{masterSR}
\eey
Derivative of the logarithm of \eq{masterSR} with respect to $-1/M$ gives the expressions for $\bar\Lambda_{B(\pm)}$,
\ben
\bar\Lambda_{B(\pm)}=\frac{\frac{\partial}{\partial(-1/M)}\int_{0}^{\omega_{\rm th}} d\omega e^{-\omega/M}\frac{1}{\pi}{\rm Im}\Pi_{B(\pm)}^{\rm OPE}}{\int_{0}^{\omega_{\rm th}} d\omega e^{-\omega/M}\frac{1}{\pi}{\rm Im}\Pi_{B(\pm)}^{\rm OPE}(\omega)}.
\label{LambdabarSR}
\een
Calculating the correlation function by OPE and substitute the results into the RHS of the above equations, we obtain the sum rules for $\bar\Lambda_{B(\pm)}$.
\section{Application to $\Lambda$ and $\Sigma$ channel}
In this section, we apply the method described in the previous section to $I=0$ ($\Lambda$) and $I=1$ ($\Sigma$) heavy baryons.
\subsection{Interpolating fields for $\Lambda$ and $\Sigma$}
\label{interpolatingfield}
The candidates of the interpolating field for $\Lambda$ are
\bey
\eta_S&\equiv&\eps_{abc}(u_aC\gamma_5d_b)h_c,
\label{etaS}\\
\eta_P&\equiv&\eps_{abc}(u_aC d_b)\gamma_5 h_c,
\label{etaP}\\
\eta_V&\equiv&\eps_{abc}(u_aC\gamma_5\gamma_\mu  d_b)\gamma^\mu h_c,
\label{etaV}
\eey
and those for $\Sigma$,
\bey
\eta_A&\equiv&\eps_{abc}(u_aC\gamma_\mu d_b)\gamma^\mu \gamma_5 h_c,
\label{etaA}\\
\eta_T&\equiv&\frac{1}{2}\eps_{abc}(u_aC\sigma_{\mu\nu} d_b)\sigma^{\mu\nu}\gamma_5 h_c,
\label{etaT}
\eey
where $u$ and $d$ are the up and down quark fields, $h$ the effective heavy quark field in the HQET, $C$ the charge conjugation operator, $\sigma_{\mu\nu}=\frac{i}{2}\left[\gamma_\mu,\gamma_\nu\right]$ and $a$, $b$ and $c$ the color indices.
The subscripts, $S$, $P$, $V$, $A$ and $T$ stand for the channel of the light diquark fields, scalar, pseudoscalar, vector, axialvector and tensor, respectively. 
The general interpolating fields for $\Lambda$ and $\Sigma$ should be given by their linear combinations,
\bey
\eta_{\Lambda}&=&t_S\eta_S+t_P\eta_P+t_V\eta_V,
\label{etaLambda}\\
\eta_\Sigma&=&t_A\eta_A+t_T\eta_T,
\label{etaSigma}
\eey
with $t_X$ $(X=S,P,V,A,T)$ being arbitrary mixing parameters.

Some important remarks are in order here.
$\eta_{S(P)}$ couples only to positive (negative) parity states, as in the form of \eq{couplingtopositive}(\eq{couplingtonegative}),
since the effective heavy quark field, $h$, which is constrained to satisfy $\vsla h=h$, is projected onto positive parity states at its rest frame $\vecv=\vec0$.
On the other hand, $\eta_V$, $\eta_{A}$ and $\eta_{T}$ can couple with both of parity states, 
which becomes obvious if we decompose them as
\bey
\eta_{V}&=&(uC\gamma_5\gamma_0  d)\gamma^0 h+(uC\gamma_5\gamma_i  d)\gamma^i h,
\label{etaVdecomp}\\
\eta_{A}&=&(uC\gamma_i  d)\gamma^i\gamma_5 h+(uC\gamma_0  d)\gamma^0\gamma_5h,
\label{etaAdecomp}\\
\eta_{T}&=&(uC\sigma_{0i}  d)\sigma^{0i}\gamma_5h+\frac{1}{2}(uC\sigma_{ij}  d)\sigma^{ij}\gamma_5h,
\label{etaTdecomp}
\eey
(Color indices are suppressed here for simplicity.)
The first term in the right hand side of each of the above equations 
couples only with positive parity state,
while the second term
only with negative parity states.
In all the previous studies on positive parity heavy baryons in QCD sum rules \cite{shuryak,grozinyakovlev,BCDN,dai,Groote1,Groote2},
the second terms of \meq{etaVdecomp}{etaTdecomp}
are excluded consistently.
However, as was pointed out above, they couple to negative parity heavy baryons and are not allowed to be excluded for the purpose of constructing the sum rules for positive and negative parity states.

Let us define the correlators of $\eta_{X}$'s,
\bey
\Pi_{XY}(\omega)&\equiv&i\int d^4x e^{i\omega v\cdot x}\bra 0|T\left[\eta_{X}(x)\bar\eta_{Y}(0)\right]|0\ket.
\label{corfuncomponent}
\eey
Then the correlation function, \eq{hbcorfun}, for $\Lambda$ and $\Sigma$ can be written in terms of $\Pi_{XY}(\omega)$ as
\bey
\Pi_{\Lambda}(\omega)&=&\sum_{X,Y=S,P,V}t_Xt_Y\Pi_{XY}(\omega),\label{Lambdacorfun}\\
\Pi_{\Sigma}(\omega)&=&\sum_{X,Y=A,T}t_Xt_Y\Pi_{XY}(\omega).\label{Sigmacorfun}
\eey

In the following two subsections, we evaluate \eq{corfuncomponent} by OPE to derive the sum rules at leading order (LO) in $\alpha_s$-expansion and those at next-to-leading order (NLO).
To this end, we write \eq{corfuncomponent} as
\bey
\Pi_{XY}(\omega)&=&\Pi^{(0)}_{XY}(\omega)+\Pi^{(1)}_{XY}(\omega)+\cdots,
\eey
where $\Pi^{(0)}_{XY}(\omega)$ and $\Pi^{(1)}_{XY}(\omega)$ denote LO and NLO contributions in $\alpha_s$-expansion,  respectively, and the ellipsis the higher order ones.
\subsection{sum rules at LO in $\alpha_s$-expansion}
\begin{figure}[t]
\includegraphics[width=5.5cm,keepaspectratio]{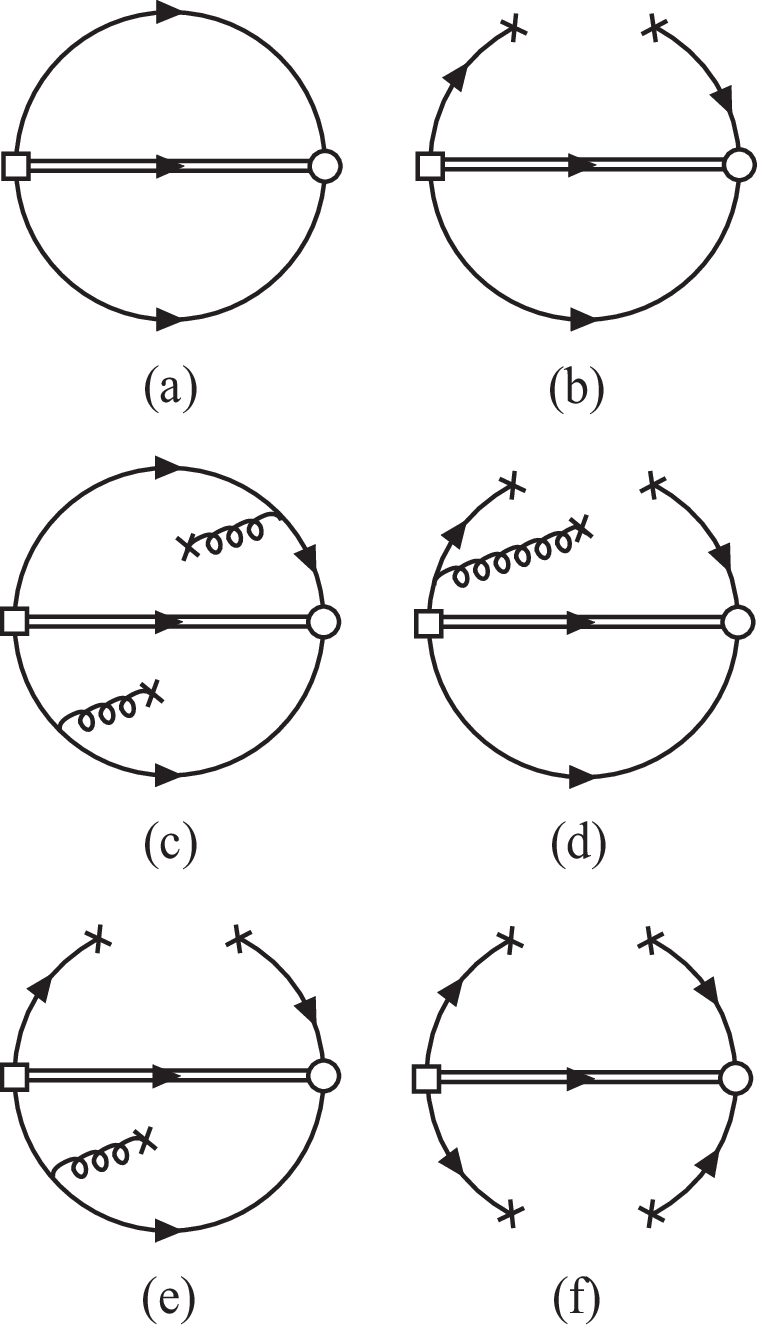}
\caption{Nonvanishing diagrams representing the terms in the OPE at LO in $\alpha_s$-expansion. (a) leading term (dimension 0), (b) dimension 3, (c) dimension 4, (d) and (e) dimension 5, (f) dimension 6 term. The single and the double lines stand for the light and the heavy quark propagators, respectively.}
\label{diagramLO}
\end{figure}
We carry out the OPE of $\Pi^{(0)}_{XY}(\omega)$. The nonvanishing contribution in the OPE up to dimension 6 operators and at LO in $\alpha_s$-expansion are represented by the Feynman diagrams shown in Fig.\ref{diagramLO}. 

First, we show the results for the components of $\Pi^{(0)}_{XY}(\omega)$ that contribute to $\Pi_{\Lambda}(\omega)$,
\bey
\Pi^{(0)}_{SS}(\omega)&=&\frac{-N_c!}{120\pi^4}\omega^5\ln(-\omega)P_+
\cr&&
+\frac{-1}{32\pi^2}\gc\omega\ln(-\omega) P_+
\cr&&
+\frac{-N_c!}{4N_c^2}\uc\dc\frac{1}{\omega}P_+
\cr&&
+\cdots,
\label{SSatLO}\\
\Pi^{(0)}_{PP}(\omega)&=&\frac{-N_c!}{120\pi^4}\omega^5\ln(-\omega)P_-
\cr&&+\frac{-1}{32\pi^2}\gc\omega\ln(-\omega) P_-
\cr&&+\frac{N_c!}{4N_c^2}\uc\dc\frac{1}{\omega}P_-
\cr&&+\cdots,
\label{PPatLO}\\
\Pi^{(0)}_{VV}(\omega)&=&\frac{-N_c!}{120\pi^4}\omega^5\ln(-\omega)(P_+ + 3P_-)
\cr&&+\frac{-1}{32\pi^2}\gc\omega\ln(-\omega)\left(P_+ - P_-\right)
\cr&&+\frac{-N_c!}{4N_c^2}\uc\dc\frac{1}{\omega}(P_+-3P_-)
\cr&&+\cdots,
\label{VVatLO}\\
\Pi^{(0)}_{SP,PS}(\omega)&=&0,
\label{PSatLO}\\
\Pi^{(0)}_{SV,VS}(\omega)&=&\frac{N_c!}{4\pi^2N_c}\left(\uc+\dc\right)\omega^{2}\ln(-\omega)P_+
\cr&&+\frac{-N_c!}{32\pi^2N_c}\left(\mixu+\mixd\right)
\cr&&\quad\times\ln(-\omega)P_+
\cr&&+\frac{1}{32\pi^2}\left(\mixu+\mixd\right)
\cr&&\quad\times\ln(-\omega)P_+
\cr&&+\cdots,
\label{SVatLO}\\
\Pi^{(0)}_{PV,VP}(\omega)&=&0,
\label{PVatLO}
\eey
where the ellipsis denotes the terms that will disappear when the Borel transformation is carried out.
Calculation of the dimension 6 (four-quark condensate) term is done by applying the factorization hypothesis.
$N_c$ is the number of colors,  $\qcon\equiv\bra 0|\q q|0\ket$, $\gc\equiv\left\bra 0|\frac{\alpha_s}{\pi}G_{\mu\nu}^aG^{a\mu\nu}|0\right\ket$, $\mixq\equiv g\left\bra0|\q\sigma_{\mu\nu}\frac{\lambda^a}{2}G^{a\mu\nu} q|0\right\ket$, where $g$ is the strong coupling constant, $\lambda^a$ the usual Gell-Mann SU(3) matrix, and $G^{a}_{\mu\nu}$ the gluon field strength.

\begin{figure}[t]
\includegraphics[width=7.5cm,keepaspectratio]{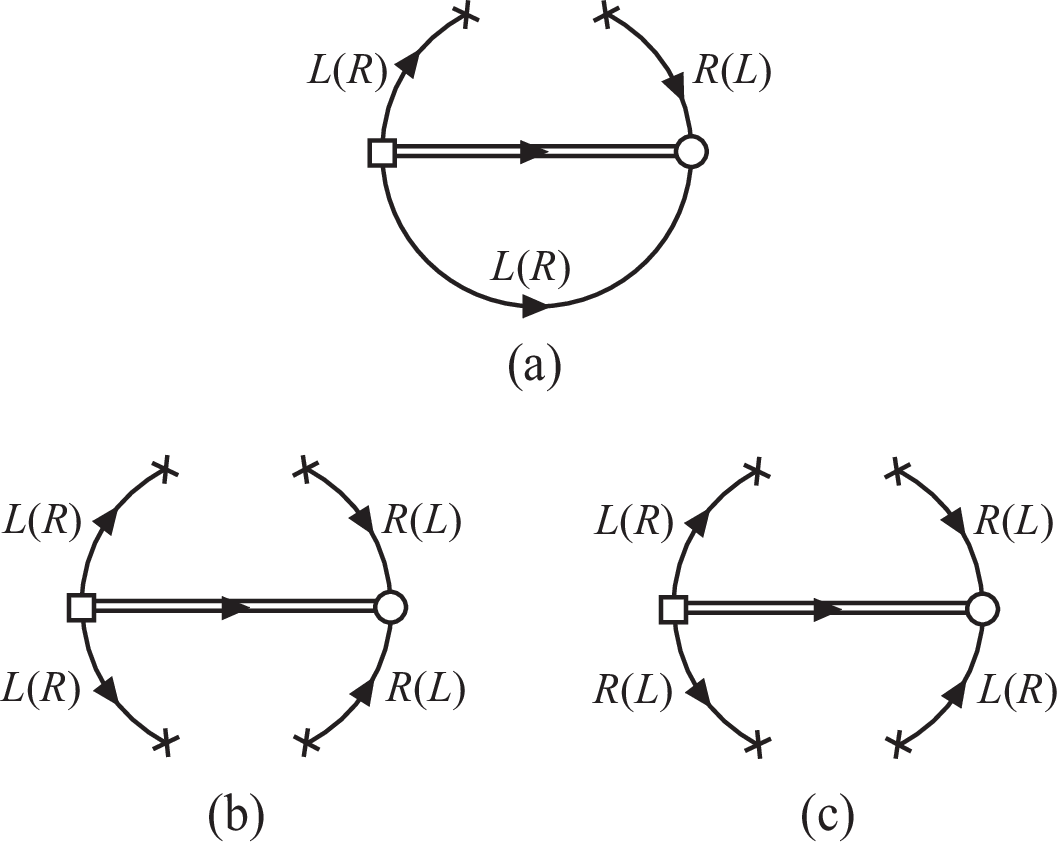}
\caption{Diagrams with explicit dependence on the chirality of light quarks, corresponding to  (a) dimension 3 term in $\Pi_{SV}$, (b) dimension 6 term in $\Pi_{SS,PP}$, and (c) that in $\Pi_{VV}$.}
\label{chiralevenorodd}
\end{figure}

Let us discuss \meq{SSatLO}{PVatLO}.
\begin{enumerate}
\renewcommand{\labelenumi}{(\roman{enumi})}
\item
$\Pi_{SS(PP)}$ has only $P_{+(-)}$ component and $\Pi_{SP,PS}=0$.
This is due to the fact that $\eta_{S(P)}$ couples only to positive (negative) parity states, as was mentioned in the previous subsection.
In contrast, $\eta_V$ can couple with both of parity states. Therefore $\Pi_{VV}$ has both of $P_{+}$ and $P_{-}$ components and $\Pi_{SV,VS}$ has only $P_{+}$.
\item
Diagonal correlators, $\Pi_{SS}$, $\Pi_{PP}$, $\Pi_{VV}$, have only chiral even condensates. On the other hand, non-diagonal correlator, $\Pi_{SV,VS}$, has only chiral odd condensates.
This can be understood by introducing right- and left-handed quark fields, $q_{R}$ and $q_{L}$ \cite{Ioffe},
\bey
\eta_S&=&(u_{R}Cd_{R}-u_{L}Cd_{L})h,
\\
\eta_P&=&(u_{R}Cd_{R}+u_{L}Cd_{L})\gamma_5 h,
\\
\eta_V&
=&(u_{R}C\gamma_\mu d_{L}-u_{L}C\gamma_\mu d_{R})\gamma^\mu h.
\eey
(Color indices are suppressed here for simplicity.) In general, in the OPE of correlation functions, $\qcon$ or $\mixq$ appear only if $q_R$ is paired with $q_L$, since
$\bar q q=\bar q_{R}q_{L}+\bar q_{L}q_{R}$, $\bar q \sigma\cdot G q=\bar q_R \sigma\cdot G q_L+\bar q_L \sigma\cdot G q_R$.
In $\Pi_{SV,VS}$ this is possible if one $q_{R(L)}$ in $\eta_S$ is paired with one $q_{L(R)}$ in $\eta_V$, as shown in Fig.\ref{chiralevenorodd} (a).
In $\Pi_{SS}$, $\Pi_{PP}$ and $\Pi_{VV}$, this is possible only if two pairs, i.e. $u_{R}$ and $u_{L}$,
$d_{R}$ and $d_{L}$, are formed at once, yielding the term $\bra 0|\u u\d d|0\ket$ (see Fig.\ref{chiralevenorodd} (b) and (c)).
\item
$\Pi_{PV,VP}$ is vanishing, although one may think that $\Pi_{PV,VP}$ has $P_{-}$ component, since both of $\eta_P$ and $\eta_V$ can couple with negative parity states, and have chiral odd condensates if one $q_{R(L)}$ in $\eta_P$ is paired with $q_{L(R)}$ in $\eta_V$.
In reality, however, the pair yields the term $\bra \bar q\gamma_5 q\ket$, which is vanishing.
\end{enumerate}

Next, we show the OPE results for $\Pi^{(0)}_{XY}(\omega)$ related with $\Pi_{\Sigma}(\omega)$,
\bey
\Pi^{(0)}_{AA}(\omega)&=&\frac{-N_c!}{120\pi^4}\omega^5\ln(-\omega)(3P_+ + P_-)
\cr&&+\frac{1}{32\pi^2}\gc\omega\ln(-\omega)\left(P_+ - P_- \right)
\cr&&+\frac{-N_c!}{4N_c^2}\uc\dc\frac{1}{\omega}(3P_+ - P_-)
\cr&&+\cdots,
\label{AAatLO}\\
\Pi^{(0)}_{TT}(\omega)&=&\frac{-N_c!}{40\pi^4}\omega^5\ln(-\omega)(P_+ + P_-)
\cr&&+\frac{1}{32\pi^2}\gc\omega\ln(-\omega)\left(P_+ + P_-\right)
\cr&&+\frac{-3N_c!}{4N_c^2}\uc\dc\frac{1}{\omega}(P_+ - P_-)
\cr&&+\cdots,
\label{TTatLO}\\
\Pi^{(0)}_{AT,TA}(\omega)&=&\frac{3N_c!}{4\pi^2N_c}\left(\uc+\dc\right)\omega^2\ln(-\omega)P_+
\cr&&+\frac{-3N_c!}{32\pi^2N_c}\left(\mixu+\mixd\right)
\cr&&\quad\times\ln(-\omega)P_+
\cr&&+\frac{-1}{32\pi^2}\left(\mixu+\mixd\right)
\cr&&\quad\times\ln(-\omega)P_+
\cr&&+\cdots.
\label{ATatLO}
\eey
Again the ellipsis denotes the terms that will vanish after the Borel transformation is applied.

In order to discuss \meq{AAatLO}{ATatLO}, we decompose $\eta_{A}$ as
\ben
\eta_{A}=\eta_{\overline A}+\eta_{\underline A},
\een
where
\ben
\eta_{\overline A}=(uC\gamma_i  d)\gamma^i\gamma_5 h,\quad \eta_{\underline A}=(uC\gamma_0  d)\gamma^0\gamma_5h.
\een
Then $\Pi_{AA}$ and $\Pi_{AT,TA}$ can also be decomposed as
\bey
\Pi_{AA}(\omega)&=&\Pi_{\overline A\overline A}(\omega)+\Pi_{\underline A\underline A}(\omega)
\cr&&+\Pi_{\overline A\underline A}(\omega)+\Pi_{\underline A\overline A}(\omega),
\label{decompAA}\\
\Pi_{AT,TA}(\omega)&=&\Pi_{\overline A T,T\overline A}(\omega)+\Pi_{\underline A T,T\underline A }(\omega).
\label{decompAT}
\eey
Following \eqs{decompAA}{decompAT}, one can decompose \eqs{AAatLO}{ATatLO} as
\bey
\Pi^{(0)}_{\overline A\overline A}(\omega)&=&\frac{-N_c!}{120\pi^4}\omega^5\ln(-\omega)\cdot 3P_+
\cr&&+\frac{1}{32\pi^2}\gc\omega\ln(-\omega)P_+
\cr&&+\frac{-N_c!}{4N_c^2}\uc\dc\frac{1}{\omega}\cdot 3P_+
\cr&&+\cdots,
\label{AbarAbaratLO}
\\
\Pi^{(0)}_{\underline A\underline A}(\omega)&=&\frac{-N_c!}{120\pi^4}\omega^5\ln(-\omega)P_-
\cr&&+\frac{-1}{32\pi^2}\gc\omega\ln(-\omega)P_-
\cr&&+\frac{N_c!}{4N_c^2}\uc\dc\frac{1}{\omega}P_- 
\cr&&+\cdots,
\label{AubarAubaratLO}
\\
\Pi^{(0)}_{\overline A\underline A,\underline A\overline A}(\omega)&=&0,
\label{AbarAubaratLO}\\
\Pi^{(0)}_{\overline A T,T\overline A}(\omega)&=&\frac{3N_c!}{4\pi^2N_c}\left(\uc+\dc\right)\omega^2\ln(-\omega)P_+
\cr&&+\frac{-3N_c!}{32\pi^2N_c}\left(\mixu+\mixd\right)
\cr&&\quad\times\ln(-\omega)P_+
\cr&&+\frac{-1}{32\pi^2}\left(\mixu+\mixd\right)
\cr&&\quad\times\ln(-\omega)P_+
\cr&&+\cdots,
\label{AbarTatLO}
\\
\Pi^{(0)}_{\underline A T,T\underline A}(\omega)&=&0.
\label{AubarTatLO}
\eey
Using \meq{AbarAbaratLO}{AubarTatLO} instead of \eqs{AAatLO}{ATatLO}, one can interpret \meq{AAatLO}{ATatLO} completely in parallel with \meq{SSatLO}{PVatLO}.
\begin{enumerate}
\renewcommand{\labelenumi}{(\roman{enumi})}
\item
$\Pi_{\overline A\overline A(\underline A\underline A)}$ has only $P_{+(-)}$ component and $\Pi_{\overline A\underline A,\underline A\overline A}=0$.
This is due to the fact that $\eta_{\overline A(\underline A)}$ couples only to positive (negative) parity states.
On the other hand, $\eta_T$ can couple with both of parity states. Therefore $\Pi_{TT}$ has both of $P_{+}$ and $P_{-}$ components and $\Pi_{\overline A T,T\overline A}$ has only $P_{+}$.
\item
Diagonal correlators, $\Pi_{\overline A\overline A}$, $\Pi_{\underline A\underline A}$ and $\Pi_{TT}$, have only chiral even condensates, while non-diagonal correlator, $\Pi_{\overline A T,T\overline A}$, has only chiral odd condensates.
To understand this, we write $\eta_A$ and $\eta_T$ in terms of right- and left-handed quark fields,
\bey
\eta_{\overline A}&=&(u_{R}C\gamma_i d_{L}+u_{L}C\gamma_i d_{R})\gamma^i \gamma_5 h,
\\
\eta_{\underline A}&=&(u_{R}C\gamma_0 d_{L}+u_{L}C\gamma_0 d_{R})\gamma^0 \gamma_5 h,
\\
\eta_T&=&\frac{1}{2}(u_RC\sigma_{\mu\nu} d_R+u_LC\sigma_{\mu\nu} d_L)\sigma^{\mu\nu}\gamma_5 h.
\eey
In $\Pi_{\overline A T,T\overline A}$, $\qcon$ or $\mixq$ can appear if one $q_{R(L)}$ in $\eta_{\overline A}$ is paired with one $q_{L(R)}$ in $\eta_T$.
In $\Pi_{\overline A\overline A}$, $\Pi_{\underline A\underline A}$ and $\Pi_{TT}$, this can occur only if $u_{R}$ is paired with $u_{L}$ and simultaneously $d_{R}$ is paired with $d_{L}$, yielding the term $\bra 0|\u u\d d|0\ket$.
\item
$\Pi_{\underline A T,T\underline A}$ is vanishing, although it may be possible that $\Pi_{\underline A T,T\underline A}$ has $P_{-}$ component, since both of $\eta_{\underline A}$ and $\eta_T$ can couple with negative parity states, and have chiral odd condensates if one $q_{R(L)}$ in $\eta_{\underline A}$ is paired with $q_{L(R)}$ in $\eta_T$.
But in fact, the pair yields the term $\bra \bar q\gamma_5 q\ket$, which is vanishing.
\end{enumerate}

Now we have all the ingredients to derive the sum rules. Substituting \eq{Lambdacorfun} in which the OPE results \meq{SSatLO}{PVatLO} are used as $\Pi_{XY}(\omega)$, and \eq{Sigmacorfun} in which \meq{AAatLO}{ATatLO} are used, into the RHS of 
\eq{LambdabarSR}, we obtain the QCD sum rules for $\bar\Lambda_{\Lambda(\pm)}$ and $\bar\Lambda_{\Sigma(\pm)}$ at LO in $\alpha_s$-expansion,
\ben
\bar\Lambda_{B(\pm)}=\frac{\frac{\partial}{\partial(-1/M)}P_{B(\pm)}^{(0)}(M)}{P_{B(\pm)}^{(0)}(M)},\quad (B=\Lambda,\Sigma),
\label{LOSR}
\een
where $P_{B(\pm)}^{(0)}(M)$ are given by
\bey
P_{\Lambda(+)}^{(0)}(M)&=&\frac{t_S^2+t_V^2}{20\pi^4}E_5(M,\omega_{th})
\cr&&-\frac{t_St_V}{\pi^2}\left(\uc+\dc\right)E_2(M,\omega_{th})
\cr&&+\frac{t_S^2+t_V^2}{32\pi^2}\gc E_1(M,\omega_{th})
\cr&&
+\frac{t_St_V}{16\pi^2}\left(\mixu+\mixd\right)
\cr&&\quad\times E_0(M,\omega_{th})
\cr&&+\frac{t_S^2+t_V^2}{6}\uc\dc,
\label{PLOLambda+}\\
P_{\Lambda(-)}^{(0)}(M)&=&\frac{t_P^2+3t_V^2}{20\pi^4}E_5(M,\omega_{th})
\cr&&+\frac{t_P^2-t_V^2}{32\pi^2}\gc E_1(M,\omega_{th})
\cr&&-\frac{t_P^2+3t_V^2}{6}\uc\dc,
\label{PLOLambda-}\\
P_{\Sigma(+)}^{(0)}(M)&=&\frac{3(t_A^2+t_T^2)}{20\pi^4}E_5(M,\omega_{th})
\cr&&-\frac{3t_At_T}{\pi^2}\left(\uc+\dc\right)E_2(M,\omega_{th})
\cr&&-\frac{t_A^2+t_T^2}{32\pi^2}\gc E_1(M,\omega_{th})
\cr&&+\frac{7t_At_T}{16\pi^2}\left(\mixu+\mixd\right)
\cr&&\quad\times E_0(M,\omega_{th})
\cr&&+\frac{t_A^2+t_T^2}{2}\uc\dc,
\label{PLOSigma+}\\
P_{\Sigma(-)}^{(0)}(M)&=&\frac{t_A^2+3t_T^2}{20\pi^4}E_5(M,\omega_{th})
\cr&&+\frac{t_A^2-t_T^2}{32\pi^2}\gc E_1(M,\omega_{th})
\cr&&-\frac{t_A^2+3t_T^2}{6}\uc\dc.
\label{PLOSigma-}
\eey
In the above equations, 
\ben
E_n(M,\omega)\equiv\int_{0}^{\omega} d\omega' \omega'^n e^{-\omega'/M},
\een
and the explicit value of the number of colors $N_c=3$ is substituted.
The numerator in \eq{LOSR} is easily obtained by using the property,
\ben
\frac{\partial}{\partial(-1/M)}E_n(M,\omega)=E_{n+1}(M,\omega).
\een

A noticeable feature of the above sum rules is that the sum rules for negative parity states do not have chiral odd terms ($\qcon$, $\mixq$, \ldots).
As was pointed out in the discussion about \meq{SSatLO}{PVatLO} and \meq{AAatLO}{ATatLO}, chiral odd condensates contribute only through the non-diagonal correlators, $\Pi_{SV}$ and $\Pi_{AT}$.
However, in the case of heavy baryons, $\Pi_{SV}$ and $\Pi_{AT}$ have only positive parity component. As a result, only the positive parity sum rules receive the contribution of chiral odd condensates but the negative parity sum rules not.
This feature of the sum rules for heavy baryons is in contrast to those for nucleons and hyperons, where chiral odd condensates contribute to both of parity states but with opposite sign, which increase the mass of negative parity and make the mass difference between positive and negative parity large \cite{JKO,JO,KMN1,KMN2}.

Also note that while the general interpolating field of $\Lambda$, \eq{etaLambda}, has three mixing parameters, $t_P$, $t_S$ and $t_V$, the sum rule for $\bar\Lambda_{\Lambda(+)}$ depends on two of them, $t_{S}$ and $t_V$, and that for $\bar\Lambda_{\Lambda(-)}$ on $t_{P}$ and $t_V$. This is due to the fact that $\eta_{P(S)}$ in \eq{etaLambda}, whose coefficient is $t_{P(S)}$, does not couple to positive (negative) parity states. 
As a result, $\bar\Lambda_{\Lambda(+)}$ depends on the mixing parameters via their ratio, $t_V/t_{S}$, and $\bar\Lambda_{\Lambda(-)}$ via $t_V/t_{P}$. 

It is interesting that \eq{PLOLambda-} and \eq{PLOSigma-} coincides with each other, which means that the mass difference between $\Lambda(-)$ and $\Sigma(-)$ is not given within this calculation. The difference is produced by including the $\alpha_s$-correction (see \eq{PNLOLambda-} and \eq{PNLOSigma-}).
\subsection{sum rules at NLO in $\alpha_s$-expansion}
\begin{figure}[t]
\includegraphics[width=5.5cm,keepaspectratio]{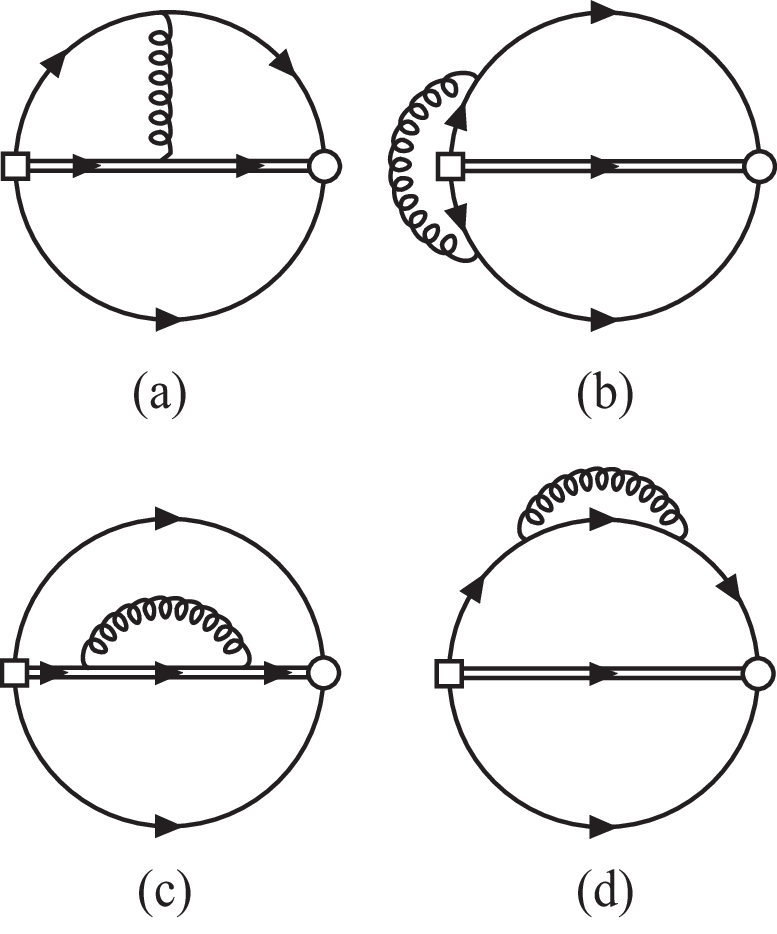}
\caption{Diagrams corresponding to NLO correction in $\alpha_s$-expansion to dimmension 0 term.}
\label{1loopcorrectiontodim0}
\end{figure}
Let us next consider NLO correction in $\alpha_s$-expansion.
We take into account the correction to the leading dimension (dimension 0) and the next dimension (dimension 3) terms in the OPE.  
For convenience, we write the dimension 0 and 3 terms in $\Pi_{XY}(\omega)$ respectively as
\bey
\Pi_{XY}(\omega)_{0}&=&\Pi^{(0)}_{XY}(\omega)_{0}+\Pi^{(1)}_{XY}(\omega)_{0}+\cdots,\\
\Pi_{XY}(\omega)_{3}&=&\Pi^{(0)}_{XY}(\omega)_{3}+\Pi^{(1)}_{XY}(\omega)_{3}+\cdots,
\eey
where $\Pi^{(0)}_{XY}(\omega)_{0,3}$ denotes LO contribution in $\alpha_s$-expansion, 
$\Pi^{(1)}_{XY}(\omega)_{0,3}$ NLO contribution, and the ellipsis the higher order.

First, let us calculate the correction to the dimension 0 term, $\Pi^{(1)}_{XY}(\omega)_0$. 
Diagrams corresponding to $\Pi^{(1)}_{XY}(\omega)_0$ are shown in Fig.\ref{1loopcorrectiontodim0}. 
The calculation of those diagrams in $D=4+2\epsilon$ space-time dimension can be performed by using the method described in \cite{broadgrozin,grozintext}. The results are presented in \meq{fig3a}{fig3d} and \meq{fig3bSP}{fig3bT}.
Expanding them in powers of $1/\epsilon$, we obtain for each component of $\Pi^{(1)}_{XY}(\omega)_0$,
\bey
\Pi^{(1)}_{SS,PP}(\omega)_0&=&
\frac{\alpha_s}{40\pi^5}N_cC_F\left[\omega^5 L\frac{1}{\epsilon}\right.
\cr&&
\left.+3\omega^5 L\left(L-\frac{2\pi^2}{27}-\frac{22}{5}\right)+\cdots
\right]P_{+,-},
\cr&&
\\
\Pi^{(1)}_{VV,AA}(\omega)_0&=&
\frac{\alpha_s}{80\pi^5}N_cC_F\left[\omega^5 L\frac{1}{\epsilon}\right.
\cr&&\left.
+3\omega^5 L\left(L-\frac{4\pi^2}{27}-\frac{223}{45}\right)+\cdots
\right]P_{+,-}
\cr&&+
\frac{3\alpha_s}{80\pi^5}N_cC_F\left[\omega^5 L\frac{1}{\epsilon}\right.
\cr&&\left.
+3\omega^5 L\left(L-\frac{4\pi^2}{27}-\frac{71}{15}\right)+\cdots
\right]P_{-,+},
\cr&&
\\
\Pi^{(1)}_{TT}(\omega)_0
&=&
\frac{\alpha_s}{40\pi^5}N_cC_F\left[\omega^5 L\frac{1}{\epsilon}\right.
\cr&&\left.
+3\omega^5 L\left(L-\frac{2\pi^2}{9}-\frac{248}{45}\right)+\cdots
\right]P_{+}
\cr&&+
\frac{\alpha_s}{40\pi^5}N_cC_F\left[\omega^5 L\frac{1}{\epsilon}\right.
\cr&&\left.
+3\omega^5 L\left(L-\frac{2\pi^2}{9}-\frac{263}{45}\right)+\cdots
\right]P_{-},
\cr&&
\eey
where $C_F=(N_c^2-1)/(2N_c)$, $L\equiv \ln(-\omega/\mu_{\rm MS})+\gamma_E/2-(1/2)\ln\pi$
with $\gamma_E$ the Euler constant,
$\mu_{\rm MS}$ is the renormalization scale in the MS scheme and the ellipsis denotes the terms that will vanish after the Borel transformation is carried out.
$L$ can be written in the $\rm\overline{MS}$ scheme as $L=\ln\left(-2\omega/\mu_{\rm\overline{MS}}\right)$ with the renormalization scale in the $\rm\overline{MS}$ scheme, $\mu_{\rm\overline{MS}}$. Later calculation is done in the $\rm\overline{MS}$ scheme and we denote $\mu_{\rm\overline{MS}}$ simply by $\mu$.

$\Pi_{XX}(\omega)_0$ is renormalized by the renormalization constant $Z_{\eta_X}$ of the interpolating field $\eta_X$ \cite{Groote1} as
\bey
\Pi_{XX}(\omega)_0&=&Z_{\eta_X}^2\Pi^{ren}_{XX}(\omega)_0.
\label{dim0renomalization}
\eey
$Z_{\eta_X}$ are given by (see Appendix \ref{etarenormalization} for derivation)
\bey
Z_{\eta_{S,P}}&=&1-\frac{3}{4}\left(1+\frac{1}{N_c}\right)\frac{\alpha_s}{\pi}\frac{1}{\epsilon}
,
\label{ZetaSP}\\
Z_{\eta_{V,A}}&=&1-\frac{3}{8}\left(1+\frac{1}{N_c}\right)\frac{\alpha_s}{\pi}\frac{1}{\epsilon}
,
\label{ZetaVA}\\
Z_{\eta_{T}}&=&1-\frac{1}{4}\left(1+\frac{1}{N_c}\right)\frac{\alpha_s}{\pi}\frac{1}{\epsilon}
.
\label{ZetaT}
\eey
By rewriting \eq{dim0renomalization} as
\bey
\Pi^{ren}_{XX}(\omega)_0&=&\Pi^{(0)}_{XX}(\omega)_0+\Pi^{(1)}_{XX}(\omega)_0
\cr&&
+\left(\frac{1}{Z_{\eta_X}^2}-1\right)\Pi^{(0)}_{XX}(\omega)_0,
\label{polecancellationindim0}
\eey
one can confirm that the pole in $\Pi^{(1)}_{XX}(\omega)_0$ (the second term in \eq{polecancellationindim0}) is cancelled by the third term (counter term). As a result, we obtain the renormalized dimension 0 term,
\bey
\Pi^{ren}_{SS,PP}(\omega)_0&=&
\left[
-\frac{1}{20\pi^4}\omega^5 L\right.
\cr&&\left.
+\frac{3\alpha_s}{10\pi^5}\omega^5 L
\left(L-\frac{2}{27}\pi^2-\frac{22}{5}\right)\right]P_{+,-},
\cr&&
\label{SSPPd0atNLO}\\
\Pi^{ren}_{VV,AA}(\omega)_0&=&
\left[-\frac{1}{20\pi^4}\omega^5 L\right.
\cr&&\left.+\frac{3\alpha_s}{20\pi^5}\omega^5 L
\left(L-\frac{4}{27}\pi^2-\frac{223}{45}\right)\right]P_{+,-}
\cr&&
+\left[-\frac{3}{20\pi^4}\omega^5 L\right.
\cr&&\left.+\frac{9\alpha_s}{20\pi^5}\omega^5 L\left(
L-\frac{4}{27}\pi^2-\frac{71}{15}\right)\right]
P_{-,+},
\label{VVAAd0atNLO}
\cr&&\\
\Pi^{ren}_{TT}(\omega)_0&=&
\left[-\frac{3}{20\pi^4}\omega^5 L\right.
\cr&&\left.+\frac{3\alpha_s}{10\pi^5}\omega^5 L\left(L-\frac{2}{9}\pi^2-\frac{248}{45}\right)\right]P_{+}
\cr&&
+\left[-\frac{3}{20\pi^4}\omega^5 L\right.
\cr&&\left.+\frac{3\alpha_s}{10\pi^5}\omega^5 L\left(L-\frac{2}{9}\pi^2-\frac{263}{45}\right)\right]P_{-},
\cr&&
\label{TTd0atNLO}
\eey
where the number of colors is fixed to be $N_c=3$.
\begin{figure}[t]
\includegraphics[width=5.5cm,keepaspectratio]{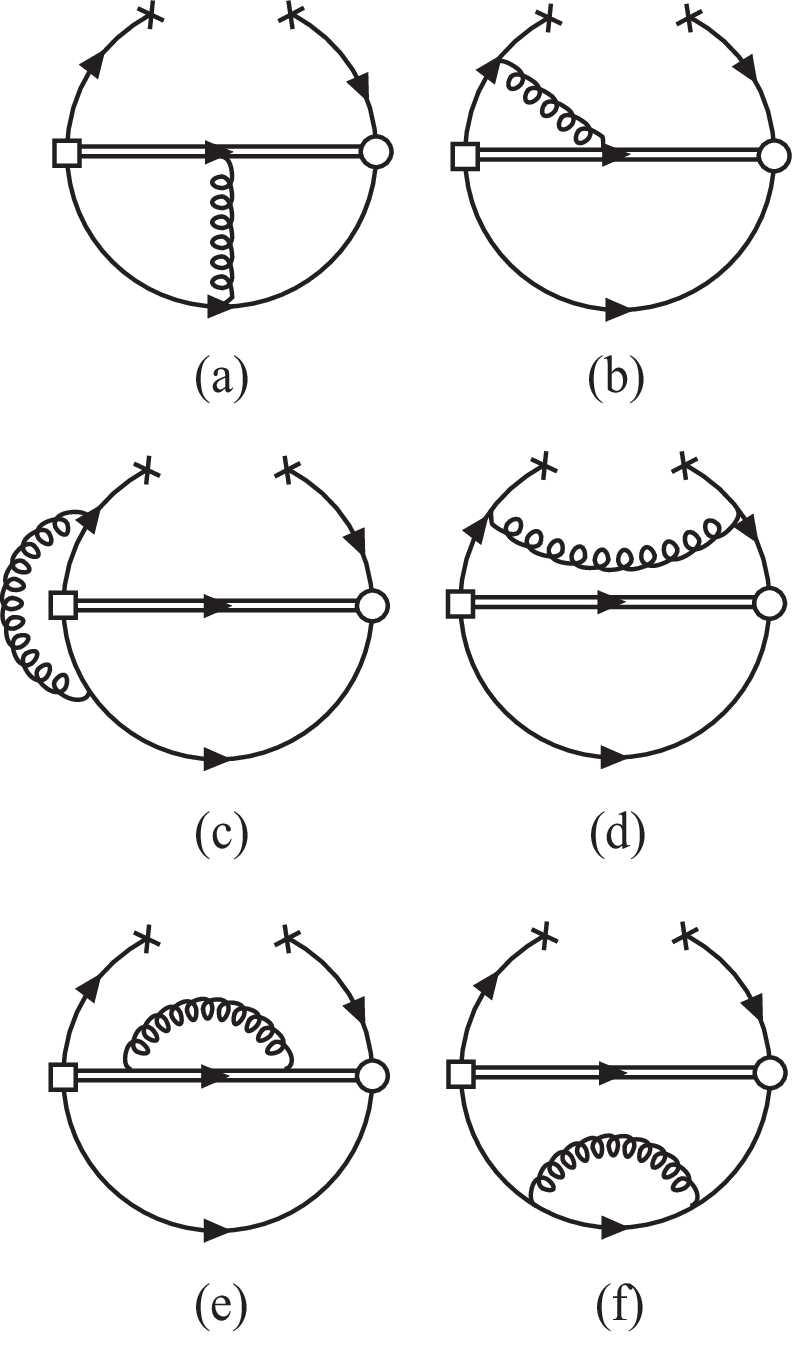}
\caption{Diagrams corresponding to NLO correction in $\alpha_s$-expansion to dimmension 3 term.}
\label{1loopcorrectiontodim3}
\end{figure}

Next, we calculate the correction to the dimension 3 term. 
Diagrams corresponding to $\Pi^{(1)}_{XY}(\omega)_3$ are shown in Fig.\ref{1loopcorrectiontodim3}.
The results of the calculation are given in \meq{fig4a}{fig4f}.
Expansion of them in powers of $1/\epsilon$ yields
\bey
\Pi^{(1)}_{SV,VS}(\omega)_3&=&
-\frac{3\alpha_s}{16\pi^3}C_F\left[\omega^2 L\frac{1}{\epsilon}\right.
\cr&&\left.
+2\omega^2 L\left(L-\frac{2\pi^2}{9}-\frac{29}{6}\right)+\cdots
\right]
\cr&&\times\left(\uc+\dc\right)P_{+},
\\
\Pi^{(1)}_{AT,TA}(\omega)_3&=&
\frac{3\alpha_s}{16\pi^3}C_F\left[\omega^2 L\frac{1}{\epsilon}\right.
\cr&&\left.
+2\omega^2 L\left(L+\frac{2\pi^2}{3}
+\frac{29}{6}\right)+\cdots
\right]
\cr&&\times\left(\uc+\dc\right)P_{+},
\eey
where the ellipsis again denotes the terms that will disappear when the Borel transformation is applied.
$\Pi_{XY}(\omega)_3$ is renormalized by $Z_{\eta_{X}}$ and $Z_{\bar q q}$ \cite{Groote2},
\bey
\Pi_{SV,VS}(\omega)_3&=&\frac{Z_{\eta_S}Z_{\eta_V}}{Z_{\bar{q}q}}\Pi_{SV,VS}^{ren}(\omega)_3,\\
\Pi_{AT,TA}(\omega)_3&=&\frac{Z_{\eta_A}Z_{\eta_T}}{Z_{\bar{q}q}}\Pi_{AT,TA}^{ren}(\omega)_3.
\eey
$Z_{\bar q q}$ is the renormalization constant of ${\bar q}q$ (see Appendix \ref{qcrenormalization}),
\bey
Z_{\bar{q}q}=1-3C_F\frac{\alpha_s}{4\pi\epsilon}.
\label{Zqbarq}
\eey
Multiplication of the renormalization constants cancels the pole in $\Pi^{(1)}_{XY}(\omega)_3$, providing the renormalized dimension 3 term,
\bey
\Pi_{SV,VS}^{ren}(\omega)_3
&=&\left[\frac{1}{2\pi^2}\omega^2 L-\frac{\alpha_s}{2\pi^3}\omega^2 L\left(L-\frac{2\pi^2}{9}-\frac{29}{6}\right)\right]
\cr&&\times\left(\uc+\dc\right)P_{+},
\label{SVd3atNLO}
\\
\Pi_{AT,TA}^{ren}(\omega)_3&=&\left[\frac{3}{2\pi^2}\omega^2 L
+\frac{\alpha_s}{2\pi^3}\omega^2 L\left(L+\frac{2\pi^2}{3}
+\frac{29}{6}\right)\right]
\cr&&\times\left(\uc+\dc\right)P_{+},
\label{ATd3atNLO}
\eey
where the number of colors is fixed to be $N_c=3$.

Now we can obtain the sum rules for $\bar\Lambda_{B(\pm)}$ at NLO in $\alpha_s$-expansion.
Substituting \meq{SSatLO}{PVatLO} and \meq{AAatLO}{ATatLO} with the dimension 0 terms replaced by \meq{SSPPd0atNLO}{TTd0atNLO} and dimension 3 terms by and \eqs{SVd3atNLO}{ATd3atNLO}, into the RHS of \eq{LambdabarSR}, we obtain
\bey
\bar\Lambda_{B(\pm)}&=&\frac{\frac{\partial}{\partial(-1/M)}\left[P_{B(\pm)}^{(0)}(M)+P_{B(\pm)}^{(1)}(M)_0\right]}
{P_{B(\pm)}^{(0)}(M)+P_{B(\pm)}^{(1)}(M)},
\label{NLOSR}
\eey
where $P_{B(\pm)}^{(1)}(M)$ denote the contributions of NLO terms in $\alpha_s$ and they consist of the corrections to dimension 0 and 3 terms,
\bey
P_{B(+)}^{(1)}(M)&=&P_{B(+)}^{(1)}(M)_0+P_{B(+)}^{(1)}(M)_3,\\
P_{B(-)}^{(1)}(M)&=&P_{B(-)}^{(1)}(M)_0.
\eey
Here, $P_{B(\pm)}^{(1)}(M)_{0}$ correspond to the contributions from NLO corrections to dimension 0 terms and their explicit expressions for $\Lambda$ and $\Sigma$ are given by,
\bey
P_{\Lambda(+)}^{(1)}(M)_0
&=&
t_S^2\frac{-3\alpha_s}{10\pi^5}\bigg[2 F_5(M,\omega_{th},\mu)
\cr&&
+E_5(M,\omega_{th})\left(\ln 4-\frac{2}{27}\pi^2-\frac{22}{5}\right)\bigg]
\cr&&
+t_V^2\frac{-3\alpha_s}{20\pi^5}
\bigg[2F_5(M,\omega_{th},\mu)
\cr&&
+E_5(M,\omega_{th})\left(\ln 4-\frac{4}{27}\pi^2-\frac{223}{45}\right)\bigg],
\cr&&\label{PNLOLambda+}\\
P_{\Lambda(-)}^{(1)}(M)_0
&=&
t_P^2\frac{-3\alpha_s}{10\pi^5}
\bigg[2F_5(M,\omega_{th},\mu)
\cr&&+E_5(M,\omega_{th})\left(\ln 4-\frac{2}{27}\pi^2
-\frac{22}{5}\right)\bigg]
\cr&&
+t_V^2\frac{-9\alpha_s}{20\pi^5}
\bigg[2F_5(M,\omega_{th},\mu)
\cr&&+E_5(M,\omega_{th})\left(\ln 4-\frac{4}{27}\pi^2-\frac{71}{15}\right)\bigg],
\cr&&\label{PNLOLambda-}\\
P_{\Sigma(+)}^{(1)}(M)_0
&=&
t_A^2\frac{-9\alpha_s}{20\pi^5}
\bigg[
2F_5(M,\omega_{th},\mu)
\cr&&+E_5(M,\omega_{th})\left(\ln 4-\frac{4}{27}\pi^2-\frac{71}{15}\right)\bigg]
\cr&&
+t_T^2\frac{-3\alpha_s}{10\pi^5}\bigg[
2F_5(M,\omega_{th},\mu)
\cr&&+E_5(M,\omega_{th})\left(\ln 4-\frac{2}{9}\pi^2-\frac{248}{45}\right)\bigg],
\cr&&\label{PNLOSigma+}\\
P_{\Sigma(-)}^{(1)}(M)_0
&=&
t_A^2\frac{-3\alpha_s}{20\pi^5}\bigg[
2F_5(M,\omega_{th},\mu)
\cr&&+E_5(M,\omega_{th})\left(\ln 4-\frac{4}{27}\pi^2-\frac{223}{45}\right)\bigg]
\cr&&
+t_T^2\frac{-3\alpha_s}{10\pi^5}
\bigg[
2F_5(M,\omega_{th},\mu)
\cr&&+E_5(M,\omega_{th})\left(\ln 4-\frac{2}{9}\pi^2-\frac{263}{45}\right)\bigg],
\cr&&\label{PNLOSigma-}
\eey
and $P_{B(+)}^{(1)}(M)_{3}$ are those from the corrections to dimension 3 terms,
\bey
P_{\Lambda(+)}^{(1)}(M)_3
&=&t_St_V\frac{\alpha_s}{\pi^3}\bigg[
2 F_2(M,\omega_{th},\mu)
\cr&&+E_2(M,\omega_{th})\left(\ln 4-\frac{2}{9}\pi^2-\frac{29}{6}\right)
\bigg]
\cr&&\times\left(\uc+\dc\right),\\
P_{\Sigma(+)}^{(1)}(M)_3
&=&
t_At_T\frac{-\alpha_s}{\pi^3}\bigg[
2 F_2(M,\omega_{th},\mu)
\cr&&+E_2(M,\omega_{th})\left(\ln 4+\frac{2}{3}\pi^2+\frac{29}{6}\right)
\bigg]
\cr&&\times\left(\uc+\dc\right).
\eey
In the above equations, 
\bey
F_n(M,\omega,\mu)\equiv\int_{0}^{\omega} d\omega' \omega'^n \ln\left(\frac{\omega'}{\mu}\right) e^{-\omega'/M},
\eey
and the derivative in the numerator of \eq{NLOSR} can be obtained by using the relation,
\ben
\frac{\partial}{\partial(-1/M)}F_n(M,\omega,\mu)=F_{n+1}(M,\omega,\mu).
\een
\section{Numerical analysis}
\begin{table}
\begin{center}
\begin{tabularx}{80mm}{Xl}
\hline
condensate&value\\
\hline
$\uc=\dc\equiv\qcon$&$(-0.24\pm 0.02\,\rm GeV)^3$\\
$\gc$&$(0.012\pm 0.006)\,\rm GeV^4$\\
$\mixu=\mixd$&$(0.8\pm 0.2)\qcon$\\
\hline
\end{tabularx}
\end{center}
\caption{Values of the vacuum condensates at the normalization scale $\mu=1\,{\rm GeV}$.}
\label{condensates}
\end{table}
In this section, we give a numerical analysis of the sum rules for $\bar\Lambda$, \eqs{LOSR}{NLOSR}
in detail.

We use the standard values for vacuum condensates collected in Table \ref{condensates}.
The value of the strong coupling constant is taken to be $\alpha_s(\mu=1\,\rm GeV)=0.47$, which is consistent with the world average \cite{Bmeson3body}.

We analyze \eqs{LOSR}{NLOSR} in the following procedure.
First, we search the value of the mixing parameter of the interpolating field that makes the curve of $\bar\Lambda$ as a function of the Borel parameter $M$ stable as much as possible.
Next, for thus found optimal value of the mixing parameter, we find $\omega_{th}$ such that the region of $M$, so called \lq\lq Borel Window", opens. 
The Borel window is the region satisfying the standard criterion applied in the QCD sum rule approach: the lowest pole contribution exceeds $50\%$ in the spectral function
\ben
\frac{\int^{\omega_{th}}_{0}d\omega\rho(\omega)}{\int^{\infty}_{0}d\omega\rho(\omega)}\geq 50\%,
\een
and the magnitude of the highest order (dimension 6) term in OPE is less than $10\%$,
\ben
\frac{\int^{\infty}_{0}d\omega\rho(\omega)_{\rm dim.6}}{\int^{\infty}_{0}d\omega\rho(\omega)_{\rm full}}\leq 10\%.
\een
Finally, we examine whether the curve of $\bar\Lambda$ stabilize or not as $\omega_{th}$ is varied.
If so, the value of $\bar\Lambda$ in the stability plateau yields a prediction of the heavy baryon mass.

\subsection{$\Lambda(+)$}
In this subsection, we consider positive parity $\Lambda$.
As was mentioned in the comments on \eq{LOSR}, $\bar\Lambda$ depends on the mixing parameter $t_S$ and $t_V$ via their ratio, so we write $t_V/t_S=t$.

\begin{figure}[t]
   \includegraphics[width=70mm,keepaspectratio]{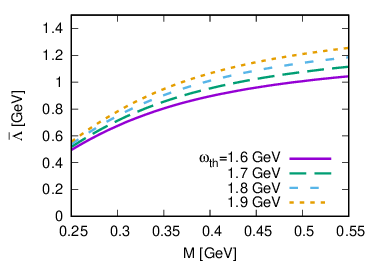}
   \caption{Borel parameter ($M$) dependence of $\bar\Lambda$ for ${\Lambda(+)}$ at LO in $\alpha_s$-expansion, \eq{LOSR}.}
   \label{Lambda+LO}
   \includegraphics[width=70mm,keepaspectratio]{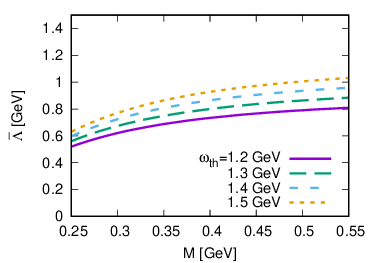}
   \caption{$M$-dependence of $\bar\Lambda$ for ${\Lambda(+)}$ at NLO in $\alpha_s$-expansion, \eq{NLOSR}.}
  \label{Lambda+NLOdim0and3}
\includegraphics[width=7cm,keepaspectratio]{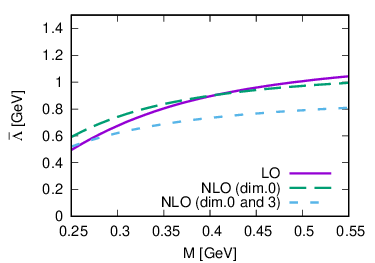}
\caption{$M$-dependence of $\bar\Lambda$ for ${\Lambda(+)}$ at LO, that with NLO correction only to dimension 0 term, and that with the corrections to dimension 0 and 3 terms. $\omega_{th}$ is taken to be 1.6GeV, 1.4GeV and 1.2GeV, respectively.}
\label{Lambda+hikaku}
\end{figure}

Let us first analyze the sum rule at LO in $\alpha_s$-expansion, \eq{LOSR}.
The curve of $\bar\Lambda$ as a function of $M$ is found to be stable for $t\simeq 0.5\sim 2.0$.
For negative values of $t$, stability is much worse. Here we fix $t=0.5$ since the results are not changed if we vary the value of $t$ in the range $t=0.5\sim2.0$.
The Borel window opens for $\omega_{\rm th}\aplg 1.6{\rm GeV}$.
Namely, if we increase $\omega_{\rm th}$, the Borel window starts to open at $\omega_{\rm th}\simeq 1.6{\rm GeV}$, and becomes wider. The lower bound of the window is $M\simeq 0.38{\rm GeV}$. 
In Fig.\ref{Lambda+LO}, the curve of $\bar\Lambda$ is plotted for several values of $\omega_{\rm th}$ in the range $\omega_{\rm th}=1.6\sim1.9{\rm GeV}$. We see that $\bar\Lambda$ does not stabilize as $\omega_{\rm th}$ is varied.

Next, we consider the sum rule at NLO in $\alpha_s$-expansion, \eq{NLOSR}.
As in LO sum rule, good stability is obtained for $t\simeq 0.5\sim 2.0$ and 
the stability is much worse for $t<0$. We fix $t=0.5$ for the same reason as in LO sum rule.
Borel window opens for $\omega_{\rm th}\aplg 1.2{\rm GeV}$. The lower bound of the window is $M\simeq 0.31{\rm GeV}$. 
$M$-dependence of $\bar\Lambda$ for $\omega_{\rm th}=1.2\sim1.5{\rm GeV}$ is plotted in Fig.\ref{Lambda+NLOdim0and3}. Although $\bar\Lambda$ does not stabilize as $\omega_{\rm th}$ is varied, the stability is slightly improved and $\bar\Lambda$ is reduced from that in LO sum rule.
Since the stability plateau does not appear, we cannot make any reliable prediction of $\bar\Lambda_{\Lambda(+)}$, and we give conservatively only the lower bound: 
\ben
\bar\Lambda_{\Lambda(+)}\aplg 0.6{\rm GeV}.
\een
The value of the lower bound is $\bar\Lambda$ at $M=0.31{\rm GeV}$ and $\omega_{\rm th}=1.2{\rm GeV}$, where the Borel window starts to show up.

For comparison, we show $M$-dependence of $\bar\Lambda$ at LO, that with NLO correction only to dimension 0 term, and that with the corrections to both of dimension 0 and 3 terms in Fig.\ref{Lambda+hikaku}. We see that the correction to dimension 3 term is more important than that to dimension 0 term.
\subsection{$\Lambda(-)$}
In this subsection, we consider negative parity $\Lambda$.
Since $\bar\Lambda$ depends on $t_P$ and $t_V$ via their ratio, we write $t_V/t_P=t$.

\begin{figure}[]
   \includegraphics[width=70mm,keepaspectratio]{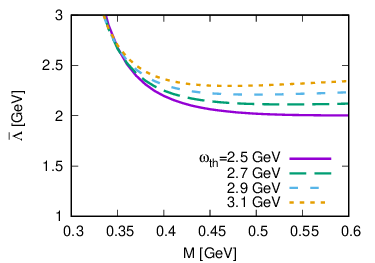}
  \caption{$M$-dependence of $\bar\Lambda$ for $\Lambda(-)$ at LO in $\alpha_s$-expansion, \eq{LOSR}.}
  \label{Lambda-LO}
   \includegraphics[width=70mm,keepaspectratio]{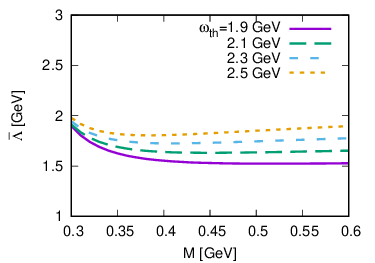}
  \caption{$M$-dependence of $\bar\Lambda$ for $\Lambda(-)$ at NLO in $\alpha_s$-expansion, \eq{NLOSR}.}
  \label{Lambda-NLO}
\includegraphics[width=7cm,keepaspectratio]{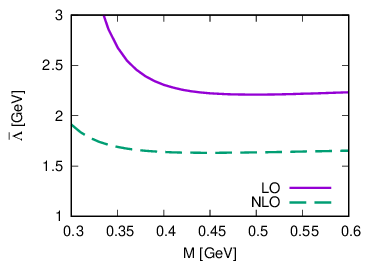}
\caption{$M$-dependence of $\bar\Lambda$ for $\Lambda(-)$ at LO and NLO. $\omega_{th}$ was taken to be 2.9GeV and 2.1GeV, respectively.}
\label{Lambda-hikaku}
\end{figure}

First we consider the sum rule at LO in $\alpha_s$, \eq{LOSR} for $\Lambda(-)$.
In this case $\bar\Lambda$ is weakly dependent on $t$, since \eq{LOSR} for $\Lambda(-)$ depends only on $t_{P}^{2}$ and $t_{V}^{2}$.
Note that \eq{LOSR} for $\Lambda(+)$ has not only $t_{S}^{2}$ and $t_{V}^{2}$ but also the cross term $t_{S}t_{V}$ and therefore $t$-dependence is stronger than that for $\Lambda(-)$.
For any value of $t$, Borel window opens for $\omega_{\rm th}\aplg 2.5{\rm GeV}$. Therefore we fix $t=0.5$.
In Fig.\ref{Lambda-LO}, we plot $\bar\Lambda$ as a function of $M$ for several values of $\omega_{\rm th}$ in the range $\omega_{\rm th}=2.5\sim 3.1{\rm GeV}$. 
We see $\bar\Lambda$ stabilizes for $\omega_{\rm th}\simeq 2.9{\rm GeV}$, and the Borel window is found to be $0.42{\rm GeV}\aplt M\aplt 0.5{\rm GeV}$.

Next we turn to NLO sum rule, \eq{NLOSR} for $\Lambda(-)$.
$t$-dependence is very weak. The reason is the same as that for LO sum rule. For any $t$, Borel window opens for $\omega_{\rm th}\aplg 1.9{\rm GeV}$. So we fix $t=0.5$.
We plot $M$-dependence of $\bar\Lambda$ in Fig.\ref{Lambda-NLO} for $\omega_{\rm th}=1.9\sim 2.5{\rm GeV}$. $\bar\Lambda$ stabilizes for $\omega_{\rm th}\simeq 2.1{\rm GeV}$, and the Borel window is found to be $0.35{\rm GeV}\aplt M\aplt 0.40{\rm GeV}$.
For this value of $\omega_{\rm th}$ and the Borel window, $\bar\Lambda_{\Lambda(-)}$ is estimated to be
\ben
\bar\Lambda_{\Lambda(-)}\simeq 1.6-1.7{\rm GeV}.
\een

In Fig.\ref{Lambda-hikaku}, we show $\bar\Lambda$ for the above two cases for comparison, from which we see the $\alpha_s$-correction significantly reduce the value of $\bar\Lambda$ and improve the stability.

\subsection{$\Sigma(+)$}
In this subsection we analyze the sum rule of $\Sigma(+)$. 
We define $t=t_T/t_A$, since $\bar\Lambda$ depends on $t_A$ and $t_T$ via their ratio.

\begin{figure}[h!]
   \includegraphics[width=70mm,keepaspectratio]{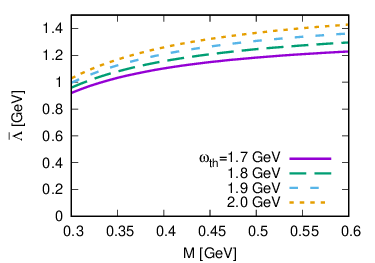}
  \caption{$M$-dependence of $\bar\Lambda$ for $\Sigma(+)$ at LO, \eq{LOSR}.}
  \label{Sigma+LO}
   \includegraphics[width=70mm,keepaspectratio]{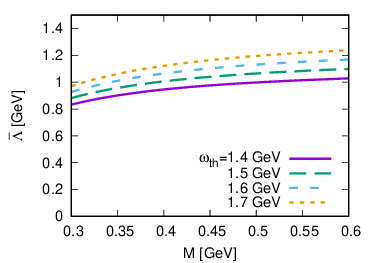}
  \caption{$M$-dependence of $\bar\Lambda$ for $\Sigma(+)$ at NLO, \eq{NLOSR}.}
  \label{Sigma+NLOdim0and3}
\includegraphics[width=7cm,keepaspectratio]{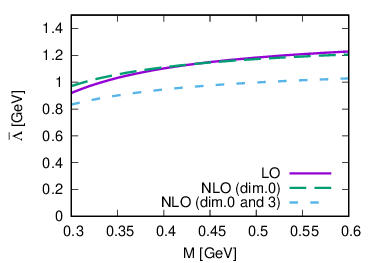}
\caption{$M$-dependence of $\bar\Lambda$ for $\Sigma(+)$ at LO in comparison with that including NLO correction only to dimension 0 term and that to dimension 0 and 3 terms. $\omega_{th}$ was taken to be 1.7GeV, 1.6GeV and 1.4GeV, respectively.}
\label{Sigma+hikaku}
\end{figure}

We begin with the sum rule at LO in $\alpha_s$, \eq{LOSR} for $\Sigma(+)$.
The best stability is obtained for $t\simeq 1.0$.
Borel window opens for $\omega_{\rm th}\aplg 1.7{\rm GeV}$. 
$M$-dependence of $\bar\Lambda$ for $\omega_{\rm th}=1.7\sim2.0{\rm GeV}$ is plotted in Fig.\ref{Sigma+LO}.

$\bar\Lambda$ at NLO in $\alpha_s$, \eq{NLOSR} for $\Sigma(+)$, is most stable for $t\simeq 1.0$.
Borel window opens for $\omega_{\rm th}\aplg 1.4{\rm GeV}$. The lower bound of the window is $M\simeq 0.33{\rm GeV}$. 
$M$-dependence of $\bar\Lambda$ is plotted in Fig.\ref{Sigma+NLOdim0and3} for $\omega_{\rm th}=1.4\sim1.7{\rm GeV}$. The curve does not stabilize as $\omega_{\rm th}$ is varied. 
As in the case of $\Lambda{(+)}$, we can determine only the lower bound of $\bar\Lambda$: 
\ben
\bar\Lambda_{\Sigma(+)}\aplg 0.9{\rm GeV}.
\een
The value of the lower bound corresponds to  $\bar\Lambda$ at $M=0.33{\rm GeV}$ and $\omega_{\rm th}=1.4{\rm GeV}$, above which the Borel window opens.

In Fig.\ref{Sigma+hikaku}, we show $M$-dependence of $\bar\Lambda$ at LO, that with NLO correction only to dimension 0 term, and that with the corrections to both of dimension 0 and 3 terms. 
We see that the stability is slightly improved by including the $\alpha_s$-correction and  the correction to dimension 3 term is more important than that to dimension 0 term.
\subsection{$\Sigma(-)$}
In this subsection, we consider the sum rule for negative parity $\Sigma$.

\begin{figure}[t]
\includegraphics[width=7cm,keepaspectratio]{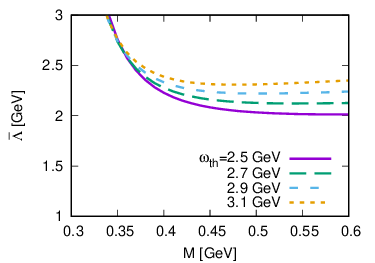}
\caption{$M$-dependence of $\bar\Lambda$ for $\Sigma(-)$ at LO, \eq{LOSR}.}
\label{Sigma-LO}
\includegraphics[width=7cm,keepaspectratio]{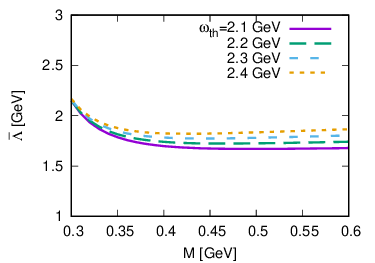}
\caption{$M$-dependence of $\bar\Lambda$ for $\Sigma(-)$ at NLO, \eq{NLOSR}.}
\label{Sigma-NLO}
\includegraphics[width=7cm,keepaspectratio]{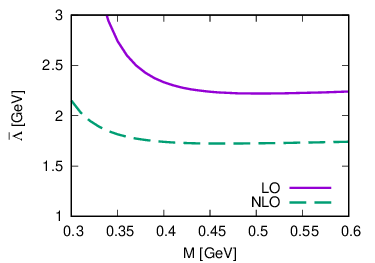}
\caption{$M$-dependence of $\bar\Lambda$ of $\Sigma(-)$ at LO and NLO. $\omega_{th}$ was taken to be 2.9GeV and 2.2GeV, respectively.}
\label{Sigma-hikaku}
\end{figure}

We begin with LO sum rule, \eq{LOSR} for $\Sigma(-)$.
The dependence on $t$ is weak, because of the absence of the cross term, $t_At_T$. For any $t$, Borel window opens for $\omega_{\rm th}\aplg 2.5{\rm GeV}$. We fix $t=1.0$.
For $\omega_{\rm th}=2.5\sim3.1{\rm GeV}$, $M$-dependence of $\bar\Lambda$ is plotted in Fig.\ref{Sigma-LO}. 

The NLO sum rule, \eq{NLOSR} for $\Sigma(-)$, is also weakly dependent on $t$. 
For any $t$, Borel window opens for $\omega_{\rm th}\aplg 2.1{\rm GeV}$. So We fix $t=1.0$.
For $\omega_{\rm th}=2.1\sim2.4{\rm GeV}$, $M$-dependence of $\bar\Lambda$ is plotted in Fig.\ref{Sigma-NLO}. 
At $\omega_{\rm th}\simeq 2.2{\rm GeV}$, $\bar\Lambda$ stabilizes and Borel window is $0.36{\rm GeV}\aplt M\aplt 0.40{\rm GeV}$.
We can make an estimate of $\bar\Lambda_{\Sigma(-)}$ from the $\bar\Lambda$ for $\omega_{\rm th}$ and the Borel window found above,
\ben
\bar\Lambda_{\Sigma(-)}\simeq 1.7-1.8{\rm GeV}.
\een

In Fig.\ref{Sigma-hikaku}, we show $\bar\Lambda$ for the above two cases for comparison, from which we see the $\alpha_s$-correction significantly reduce the value of $\bar\Lambda$ and improve the stability.

\section{Summary and discussion}
We have formulated the QCD sum rules for positive and negative parity states of heavy baryon containing one heavy quark in the framework of HQET.
Choice of the interpolating field is a crucial problem in the QCD sum rule approach.
We used the general interpolating field of positive intrinsic parity without covariant derivatives for given isospin and included the components that couple with negative parity states, which were discarded in the previous studies. 
It has been shown that by applying parity projection onto the correlation function of the interpolating fields the sum rules for respective parity states can be constructed.
The sum rule enables us to evaluate the positive or negative parity heavy baryon masses relative to the heavy quark mass, $\bar\Lambda$. 
We have applied the method to $\Lambda$ and $\Sigma$ channels.
In the OPE we have taken into account the operators up to dimension 6 and included $\alpha_s$-corrections to the terms of the leading dimension and those of the next dimension.

The effects of $\alpha_s$-correction were found to be significant, especially for negative parity states.
For positive parity states, by including $\alpha_s$-corrections the Borel parameter ($M$)-stability is slightly improved, but the improvement is not enough to give any reliable predictions of $\bar\Lambda$.
It is allowed to give only a lower bound of $\bar\Lambda$:
\ben
\bar\Lambda_{\Lambda(+)}\aplg 0.6\,{\rm GeV},\quad \bar\Lambda_{\Sigma(+)}\aplg 0.9\,{\rm GeV}.
\label{Lambdabar+}
\een
For negative parity states significant improvement of $M$-stability and large reduction of $\bar\Lambda$ were found. In fact, stability plateau appears, which allows us to give an estimation of $\bar\Lambda$:
\ben
\bar\Lambda_{\Lambda(-)}\simeq 1.6-1.7\,{\rm GeV},\quad \bar\Lambda_{\Sigma(-)}\simeq 1.7-1.8\,{\rm GeV}.
\label{Lambdabar-}
\een

The mass of the heavy baryon is given by $m_{B}=m_{Q}+\bar\Lambda$ at leading order of $1/m_{Q}$ expansion in HQET, where $m_{Q}$ is the current mass of the heavy quark $Q$. When we use the $\overline{\rm MS}$ masses of $c$ and $b$ quark \cite{pdg},
\ben
m_{c}=1.27\,{\rm GeV},\quad m_{b}=4.18\,{\rm GeV}.
\een
we obtain the lower bound of the masses of positive parity states,
\bey
&& m_{\Lambda_{c}(+)}\aplg 1.87\,{\rm GeV},
\cr&& m_{\Sigma_{c}(+)}\aplg 2.17\,{\rm GeV},
\cr&& m_{\Lambda_{b}(+)}\aplg 4.78\,{\rm GeV},
\cr&& m_{\Sigma_{b}(+)}\aplg 5.08\,{\rm GeV},
\eey
which do not contradict the masses of the observed $J^{P}=1/2^{+}$ states \cite{pdg}: $\Lambda_{c}(2286)$, $\Sigma_{c}(2455)$, $\Lambda_{b}(5620)$, $\Sigma_{b}(5811)$.
For the masses of negative parity states, we obtain,
\bey
&&m_{\Lambda_{c}(-)}\simeq 2.87-2.97\,{\rm GeV},
\cr&& m_{\Sigma_{c}(-)}\simeq 2.97-3.07\,{\rm GeV},
\cr&& m_{\Lambda_{b}(-)}\simeq 5.78-5.88\,{\rm GeV},
\cr&& m_{\Sigma_{b}(-)}\simeq 5.88-5.98\,{\rm GeV}.
\eey
Observed $J^{P}=1/2^{-}$ states of $\Lambda$ are $\Lambda_{c}(2595)$ and $\Lambda_{b}(5912)$ \cite{pdg}.
Our prediction of $m_{\Lambda_{b}(-)}$ is close to the experimental value, while that of $m_{\Lambda_{c}(-)}$ is not. The deviation may be attributed to the fact that for charmed baryons HQET is inappropriate and $1/m_Q$ correction is important.
Our predictions of $m_{\Sigma_{c}(-)}$ and $m_{\Sigma_{b}(-)}$ suggest that observed $\Sigma_{c}(2800)$ and $\Sigma_{b}(6097)$ \cite{pdg}, whose spin-parity has not yet been specified, can be the candidates of $J^{P}=1/2^{-}$ states.

A notable feature of the heavy baron sum rule, which is not seen in the light baryon sum rule, is that the sum rule for the negative parity states do not have chiral odd condensates.
Chiral odd condensates contribute only through the non-diagonal correlators, owing to the property of the light diquark.
On the other hand, since the heavy quark field couples only with positive parity states, the non-diagonal correlators have only positive parity component.
Therefore, in the heavy quark limit, it is inevitable that the negative parity states do not depend on chiral odd condensates.

If the mass of the heavy quark is large but finite and heavy quark condensation occurs, the diagonal correlators
yield heavy quark condensate $\bra \bar Q Q\ket$ terms while the non-diagonal correlators $\qcon$ terms. As a result, the chiral odd term of the correlation function reads
\ben
\Pi_{\rm chiral\,odd}\sim\bra \bar Q Q\ket(P_+-P_-)+\qcon(P_++\alpha P_-)+\cdots,
\label{Piforfinitem_Q}
\een
where the ellipsis denotes higher dimensional terms and $\alpha$ is a coefficient suppressed by powers of $\mu/ m_Q$. In the heavy quark limit, \eq{Piforfinitem_Q} is reduced to
\ben
\Pi_{\rm chiral\,odd}\sim \qcon P_++\cdots,
\een
and accordingly the chiral odd condensates do not appear in the sum rule for negative parity.
\appendix
\section{calculation of the diagrams with $\alpha_{s}$-correction (Fig.\ref{1loopcorrectiontodim0} and \ref{1loopcorrectiontodim3})}
In this appendix, we explain briefly how the calculation of the diagrams including $\alpha_{s}$-correction shown in Fig.\ref{1loopcorrectiontodim0} and \ref{1loopcorrectiontodim3} is performed and present the full results.

\begin{figure}[t]
\includegraphics[width=7.2cm,keepaspectratio]{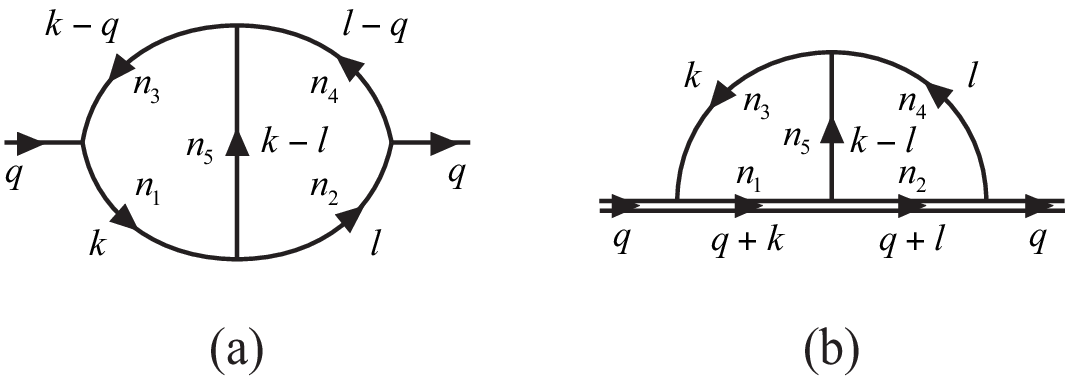}
\caption{Subdiagrams of Fig.\ref{1loopcorrectiontodim0} and \ref{1loopcorrectiontodim3}. The single (double) line stands for the massless (heavy) particle propagator.}
\label{subdiagram}
\end{figure}

Although a little bit lengthy, the calculation can be performed straightforwardly by exploiting the method invented by Grozin \cite{grozintext}.
It is convenient to calculate the diagrams in momentum space.
The diagrams in Fig.\ref{1loopcorrectiontodim0} and \ref{1loopcorrectiontodim3} can be expressed with the integrals of subdiagrms shown in Fig.\ref{subdiagram} (a) or (b). 
Fig.\ref{subdiagram} (a) is the massless particle diagram, whose explicit expressions are defined by
\bey
&&\int\frac{d^Dkd^Dl}{D_1^{n_1}D_2^{n_2}D_3^{n_3}D_4^{n_4}D_5^{n_5}}
\cr&=&-\pi^D(-q^2)^{D-\sum_in_i}G(n_1,n_2,n_3,n_4,n_5),
\cr&&
\eey
where
\bey
&&D_1=-k^2,\quad D_2=-l^2,
\cr&&D_3=-(k-q)^2,\quad D_4=-(l-q)^2,\quad D_5=-(k-l)^2.
\cr&&
\eey
Fig.\ref{subdiagram} (b) is the diagram of heavy particle defined by
\bey
&&\int\frac{d^Dkd^Dl}{D_1^{n_1}D_2^{n_2}D_3^{n_3}D_4^{n_4}D_5^{n_5}}
\cr&=&-\pi^D(-2q_0)^{2(D-n_3-n_4-n_5)}I(n_1,n_2,n_3,n_4,n_5),
\cr&&
\eey
where
\bey
&&D_1=\frac{k_0+q_0}{q_0},\quad D_2=\frac{l_0+q_0}{q_0},
\cr&& D_3=-k^2,\quad D_4=-l^2,\quad D_5=-(k-l)^2.
\eey
Therefore we first calculate those subdiagrams, namely $G$ or $I$ in the above equation, which can be expressed in terms of more easily calculable diagrams \cite{grozintext}.
Next, integrating the subdiagrams, we finally obtain the expressions of Fig.\ref{1loopcorrectiontodim0} and \ref{1loopcorrectiontodim3}.

\begin{widetext}
The results of Fig.\ref{1loopcorrectiontodim0} (a), (c) and (d) are summarized as follows,
\bey
\Pi_{XY}(\omega)_{\rm Fig.\ref{1loopcorrectiontodim0}(a)}
&=&\frac{g^2 N_cC_F}{32\pi^{3D/2}}(-\omega)^5\left(\frac{-\omega}{\mu}\right)^{3(D-4)}
\frac{\Gamma(D/2)\Gamma(7-3D)}{\Gamma(6-2D)}
\cr&&\times
\left[2I(1,1,0,1,1)-I(1,1,1,1,0)-2I(0,1,1,1,1)+I(1,1,1,1,1)\right]M_{2},
\label{fig3a}
\\
\Pi_{XY}(\omega)_{\rm Fig.\ref{1loopcorrectiontodim0}(c)}
&=&\frac{g^2 N_cC_F}{8\pi^{3D/2}}(-1)^{3D+7}(-\omega)^5\left(\frac{-\omega}{\mu}\right)^{3(D-4)}
\frac{\Gamma(D/2)^2\Gamma(7-3D)\Gamma(3-D)\Gamma(D/2-1)}{\Gamma(5-D)}M_{2},
\label{fig3c}
\\
\Pi_{XY}(\omega)_{\rm Fig.\ref{1loopcorrectiontodim0}(d)}
&=&\frac{g^2 N_cC_F}{32\pi^{3D/2}}(-1)^{2D}(-\omega)^5\left(\frac{-\omega}{\mu}\right)^{3(D-4)}
\frac{(2-D)\Gamma(D/2)\Gamma(7-3D)\Gamma(2-D/2)\Gamma(D/2-1)^2}{\Gamma(3-D/2)}M_{2},
\cr&&
\label{fig3d}
\eey
where the matrix $M_{2}$ is given by
\bey
M_{2}
=\left\{
\begin{array}{ll}
4P_\pm& (XY=SS,PP)\\
4\left[P_\pm+(D-1)P_\mp\right] & (XY=VV,AA)\\
8(D-1)\left[2P_++(D-2)P_-\right] & (XY=TT)\\
0 & (XY=\rm else)
\end{array}
\right.
\cr&&
\eey

The result of Fig.\ref{1loopcorrectiontodim0} (b) has more complicated form. For each component they are given by
\bey
\Pi_{SS,PP}(\omega)_{\rm Fig.\ref{1loopcorrectiontodim0}(b)}
&=&\frac{g^{2}N_cC_F}{64\pi^{3D/2}}(-\omega)^5\left(\frac{-\omega}{\mu}\right)^{3(D-4)}\frac{\Gamma(3D/2-3)\Gamma(7-3D)}{\Gamma(3-D)}
\cr&&\times\left\{
4\left[-G(0,1,1,1,1)-G(1,0,1,1,1)-G(1,1,0,1,1)-G(1,1,1,0,1)+G(1,1,1,1,1)\right]
\right.
\cr&&\left.
\quad-2(D-4)G(1,1,1,1,0)+2(D-2)\left[G(0,1,1,0,1)+G(1,0,0,1,1)\right]
\right\},
\label{fig3bSP}
\\
\Pi_{VV,AA}(\omega)_{\rm Fig.\ref{1loopcorrectiontodim0}(b)}
&=&\frac{g^{2}N_cC_F}{64\pi^{5D/2}}(-\omega)^5\left(\frac{-\omega}{\mu}\right)^{3(D-4)}
\frac{\Gamma(3D/2-2)\Gamma(7-3D)}{(D-1)\Gamma(3-D)}
\cr&&\times
\left\{
\left[A_{VA}+\frac{7-3D}{2(3-D)}B_{VA}\right]P_{\pm}
+(1-D)\left[A_{VA}+\frac{1}{2(3-D)}B_{VA}\right]P_\mp
\right\},
\label{fig3bVA}
\\
\Pi_{TT}(\omega)_{\rm Fig.\ref{1loopcorrectiontodim0}(b)}
&=&-\frac{g^2N_cC_F}{32\pi^{5D/2}}(-\omega)^5\left(\frac{-\omega}{\mu}\right)^{3(D-4)}\frac{\Gamma(3D/2-3)\Gamma(7-3D)}{\Gamma(3-D)}
\cr&&\times
\left\{
(1-D)\left[2A_T+\frac{8-3D}{3-D}B_T\right]P_+
+(2-D)\left[(1-D)A_T+\frac{7-4D}{3-D}B_T\right]P_-
\right\},
\label{fig3bT}
\eey
where $A_{VA}$, $B_{VA}$, $A_T$ and $B_T$ are given in terms of $G$,
\bey
A_{VA}&=&2\pi^D(D-2)[4G(-1,1,1,0,1)+2(D-5)G(0,1,1,0,1)-(D-8)G(1,1,1,1,0)
\cr&&-8G(0,1,1,1,1)+2G(1,1,1,1,-1)+2G(1,1,1,1,1)],
\\
B_{VA}&=&-2\pi^D(D-2)[4DG(-1,1,1,0,1)-8G(0,1,1,0,1)-(D-8)G(1,1,1,1,0)
\cr&&-8G(0,1,1,1,1)+2G(1,1,1,1,-1)+2G(1,1,1,1,1)],
\\
A_T&=&\frac{2\pi^D}{(D-1)(D-2)}\{
8(D-4)^2G(-1,1,1,0,1)-4(D-3)(D-4)G(1,1,1,1,-1)
\cr&&
+8(D-2)(D-3)G(0,1,1,1,1)
+2[28+2D(D-7)-(D-2)(D-4)(D-5)]G(1,0,0,1,1)
\cr&&
+[2(D-8)+(D-4)(D^2-13D+20)]G(1,1,1,1,0)
-2(D-2)(D-3)G(1,1,1,1,1)
\},
\\
B_T&=&\frac{2\pi^D}{(D-1)(D-2)}\{
4D(D-4)^2G(-1,1,1,0,1)+2(D-4)^2G(1,1,1,1,-1)
\cr&&
-8(D-2)^2G(0,1,1,1,1)
-8(D^2-6D+10)G(1,0,0,1,1)
\cr&&
-[D(D-8)^2-80]G(1,1,1,1,0)
+2(D-2)^2G(1,1,1,1,1)
\}.
\eey

The results of Fig.\ref{1loopcorrectiontodim3} are summarized as follows,
\bey
\Pi_{XY}(\omega)_{\rm Fig.\ref{1loopcorrectiontodim3}(a)}&=&\frac{-g^2C_F}{128\pi^D}(-\omega)^2\left(\frac{-\omega}{\mu}\right)^{2(D-4)}
\cr&&\times
\left[2I(1,1,0,1,1)-I(1,1,1,1,0)+2I(0,1,1,1,1)+I(1,1,1,1,1)\right]M_{{31}},
\label{fig4a}
\\
\Pi_{XY}(\omega)_{\rm Fig.\ref{1loopcorrectiontodim3}(b)}&=&\frac{-g^2C_F}{32\pi^D}(-\omega)^2\left(\frac{-\omega}{\mu}\right)^{2(D-4)}
\frac{\Gamma(D/2-1)\Gamma(D/2)\Gamma(4-D)\Gamma(6-2D)}{\Gamma(5-D)}M_{{31}},
\label{fig4b}
\\
\Pi_{XY}(\omega)_{\rm Fig.\ref{1loopcorrectiontodim3}(c)}&=&\frac{-g^2C_F}{256\pi^D}(-\omega)^2\left(\frac{-\omega}{\mu}\right)^{2(D-4)}
\frac{\Gamma(3-D/2)\Gamma(D/2-1)^2\Gamma(6-2D)}{\Gamma(4-D/2)}M_{{32}},
\label{fig4c}
\\
\Pi_{XY}(\omega)_{\rm Fig.\ref{1loopcorrectiontodim3}(d)}&=&\frac{g^2C_F(-1)^{7-D}}{64\pi^D}(-\omega)^2\left(\frac{-\omega}{\mu}\right)^{2(D-4)}
D\Gamma(D/2-2)\Gamma(D/2)\Gamma(6-2D)M_{{31}},
\label{fig4d}
\\
\Pi_{XY}(\omega)_{\rm Fig.\ref{1loopcorrectiontodim3}(e)}&=&\frac{g^2C_F}{16\pi^D}(-\omega)^2\left(\frac{-\omega}{\mu}\right)^{2(D-4)}
\frac{\Gamma(D/2-1)\Gamma(D/2)\Gamma(6-2D)\Gamma(3-D)}{\Gamma(5-D)}M_{{31}},
\label{fig4e}
\\
\Pi_{XY}(\omega)_{\rm Fig.\ref{1loopcorrectiontodim3}(f)}&=&\frac{g^2C_F(-1)^D}{64\pi^D}(-\omega)^2\left(\frac{-\omega}{\mu}\right)^{2(D-4)}
\frac{\Gamma(D/2-1)\Gamma(D/2)\Gamma(6-2D)\Gamma(2-D/2)}{\Gamma(3-D/2)}M_{{31}},
\label{fig4f}
\eey
where the matrices $M_{31}$ and $M_{32}$ are given by
\bey
M_{{31}}
&=&(\uc+\dc)\times\left\{
\begin{array}{ll}
4P_+& (XY=SV,VS)\\
8(D-1)P_+ & (XY=AT,TA)\\
0 & (XY=\rm else)
\end{array}
\right.\\
M_{{32}}&=&(\uc+\dc)\times\left\{
\begin{array}{ll}
\frac{4(D-2)(D-6)(D+1)}{D-4}P_+& (XY=SV,VS)\\
\frac{8(D-1)(D-2)(D-3)(D-6)}{D-4}P_+ & (XY=AT,TA)\\
0 & (XY=\rm else)
\end{array}
\right.
\eey
\end{widetext}
Following the description in Ref.\cite{grozintext}, the subdiagram integral, $G$ and $I$ appearing in \eq{fig3a}, \meq{fig3bSP}{fig3bT} and \eq{fig4a}, can be reduced to some combinations of the diagrams which can be easily calculated. The procedure is straightforward, but the results are so lengthy that we cannot help omitting them here.
\section{renormalization of interpolating field}
\label{etarenormalization}
In this appendix, we calculate $\alpha_{s}$-corrections of the interpolating fields, \meq{etaS}{etaT}, in $D=4+2\epsilon$ space-time dimension, and determine the corresponding renormalization constants.
\begin{figure}[t]
\includegraphics[width=6.7cm,keepaspectratio]{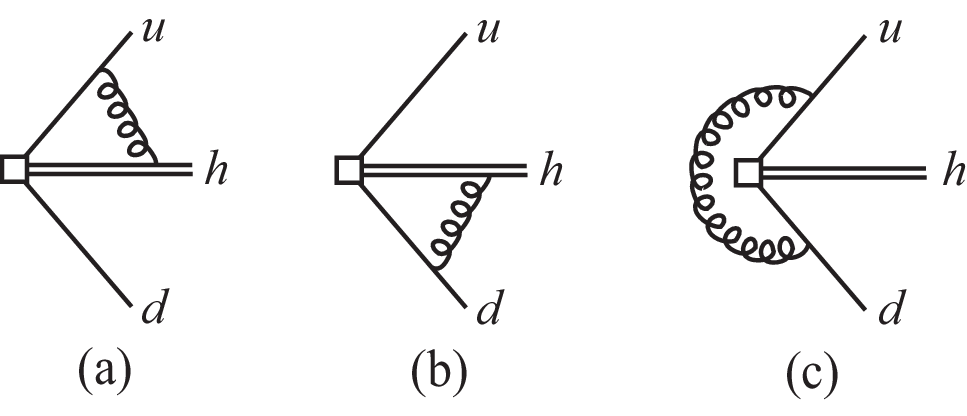}
\caption{Diagrams corresponding to $\alpha_s$-corrections to the interpolating field, \eq{etaX}.}
\label{IFcorrection}
\end{figure}

The interpolating field under consideration is
\bey
\eta_X=(qC\Gamma_X q) \Gamma'_X h,\quad (X=S,P,V,A,T),
\label{etaX}
\eey
where $\Gamma^{(')}_X$ is  
\bey
\left(\Gamma_X,\Gamma^{'}_X\right)=\left\{
\begin{array}{ll}
\left(1,\gamma_5\right)&(X=P)\\
\left(\gamma_5,1\right)&(X=S)\\
\left(\gamma_5\gamma_\mu,\gamma^\mu\right)&(X=V)\\
\left(\gamma_\mu,\gamma^\mu\gamma_5\right)&(X=A)\\
\left(\sigma_{\mu\nu},\sigma^{\mu\nu}\gamma_5\right)&(X=T)
\end{array}
\right.
\eey
Renormalization constant of the interpolating field, $Z_{\eta_X}$, is defined by
\bey
\eta_X^B=Z_{\eta_X}\eta_X^R,
\label{defZeta}
\eey
where $\eta_X^B$ and $\eta_X^R$ denote the bare and renormalized interpolating fields, respectively.
Then \eq{defZeta} can be rewritten as
\bey
\eta_X^R
&=&(q^RC\Gamma_X q^R)\Gamma'_X h^R
\cr&&+\left(\frac{Z_q\sqrt{Z_h}}{Z_{\eta_X}}-1\right)(q^RC\Gamma_X q^R)\Gamma'_X h^R
\label{etaR}
\eey
with $Z_q$ and $Z_h$ respectively being the wave function renormalization of the light quark and the heavy quark field 
given by 
\bey
Z_q=1+C_F\frac{\alpha_s}{4\pi\epsilon},\quad Z_h=1-2C_F\frac{\alpha_s}{4\pi\epsilon}.
\eey
$Z_{\eta_X}$ is determined so that the counter term (the second term in \eq{etaR}) cancels the UV-pole of the $\alpha_s$-correction of $\eta_X$.
The diagram corresponding to $\alpha_s$-correction of $\eta_X$ are depicted in Fig.\ref{IFcorrection},
and its UV-pole reads
\bey
&&\mbox{\rm UV-pole of Fig.\ref{IFcorrection}}
\cr&=&-k\left(1+\frac{1}{N_c}\right)\frac{\alpha_s}{\pi}\frac{1}{\epsilon}
\epsilon_{abc}\left(u_a^T C\Gamma_X d_b\right)\Gamma'_X  h_c,
\cr&&
\label{etaUVpole}
\eey
where the coefficient $k$ for each channel is given by 
\bey
k=\left\{
\begin{array}{l}
\frac{3}{4}\quad(X=S,P)\\
\frac{3}{8}\quad(X=V,A)\\
\frac{1}{4}\quad (X=T)
\end{array}
\right.
\eey
$Z_{\eta_X}$ is chosen so that the counter term in \eq{etaR} is equal to the minus of \eq{etaUVpole}.
$Z_{\eta_X}$ obtained in this way are \meq{ZetaSP}{ZetaT}.
\begin{figure}[]
\includegraphics[width=1.1cm,keepaspectratio]{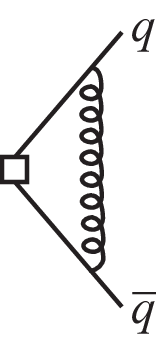}
\caption{Diagrams corresponding to $\alpha_s$-corrections to ${\bar q}q$.}
\label{qbarcorrection}
\end{figure}
\section{renormalization of $\bar q q$}
\label{qcrenormalization}
In this appendix, we determine the renormalization constant of $\bar q q$.

Renormalization constant of $\bar q q$, $Z_{\bar q q}$, is defined by
\bey
(\bar q q)^B=Z_{{\bar q}q}(\bar q q)^R,
\label{defZ3}
\eey
where $(\bar q q)^B$ and $(\bar q q)^R$ denote the bare and renormalized operators, respectively.
Then \eq{defZ3} can be rewritten as
\bey
(\bar q q)^R={\bar q}^Rq^R+\left(\frac{Z_q}{Z_{{\bar q}q}}-1\right){\bar q}^Rq^R.
\label{eta3}
\eey
On the other hand, UV-pole of $\alpha_s$-correction of $\bar q q$, shown in Fig.\ref{qbarcorrection}, reads
\bey
\mbox{\rm UV-pole of Fig.\ref{qbarcorrection}}=-C_F\frac{\alpha_s}{\pi}\frac{1}{\epsilon}{\bar q}q.
\label{qbarqUVpole}
\eey
$Z_{\bar q q}$ is determined so that the counter term (the second term in \eq{eta3}) cancels the UV-pole of the $\alpha_s$-correction of $\bar q q$, \eq{qbarqUVpole}.
Thus we obtain $Z_{\bar q q}$ as in \eq{Zqbarq}.
\clearpage
\def\Ref#1{[\ref{#1}]}
\def\Refs#1#2{[\ref{#1},\ref{#2}]}
\def\npb#1#2#3{{Nucl. Phys.\,}{\bf B{#1}},\,#2\,(#3)}
\def\npa#1#2#3{{Nucl. Phys.\,}{\bf A{#1}},\,#2\,(#3)}
\def\np#1#2#3{{Nucl. Phys.\,}{\bf{#1}},\,#2\,(#3)}
\def\plb#1#2#3{{Phys. Lett.\,}{\bf B{#1}},\,#2\,(#3)}
\def\prl#1#2#3{{Phys. Rev. Lett.\,}{\bf{#1}},\,#2\,(#3)}
\def\prd#1#2#3{{Phys. Rev.\,}{\bf D{#1}},\,#2\,(#3)}
\def\prc#1#2#3{{Phys. Rev.\,}{\bf C{#1}},\,#2\,(#3)}
\def\prb#1#2#3{{Phys. Rev.\,}{\bf B{#1}},\,#2\,(#3)}
\def\pr#1#2#3{{Phys. Rev.\,}{\bf{#1}},\,#2\,(#3)}
\def\ap#1#2#3{{Ann. Phys.\,}{\bf{#1}},\,#2\,(#3)}
\def\prep#1#2#3{{Phys. Reports\,}{\bf{#1}},\,#2\,(#3)}
\def\rpp#1#2#3{{Rept. Prog. Phys.\,}{\bf{#1}},\,#2\,(#3)}
\def\rmp#1#2#3{{Rev. Mod. Phys.\,}{\bf{#1}},\,#2\,(#3)}
\def\cmp#1#2#3{{Comm. Math. Phys.\,}{\bf{#1}},\,#2\,(#3)}
\def\ptp#1#2#3{{Prog. Theor. Phys.\,}{\bf{#1}},\,#2\,(#3)}
\def\ib#1#2#3{{\it ibid.\,}{\bf{#1}},\,#2\,(#3)}
\def\zsc#1#2#3{{Z. Phys. \,}{\bf C{#1}},\,#2\,(#3)}
\def\zsa#1#2#3{{Z. Phys. \,}{\bf A{#1}},\,#2\,(#3)}
\def\intj#1#2#3{{Int. J. Mod. Phys.\,}{\bf A{#1}},\,#2\,(#3)}
\def\sjnp#1#2#3{{Sov. J. Nucl. Phys.\,}{\bf #1},\,#2\,(#3)}
\def\pan#1#2#3{{Phys. Atom. Nucl.\,}{\bf #1},\,#2\,(#3)}
\def\app#1#2#3{{Acta. Phys. Pol.\,}{\bf #1},\,#2\,(#3)}
\def\jmp#1#2#3{{J. Math. Phys.\,}{\bf {#1}},\,#2\,(#3)}
\def\cp#1#2#3{{Coll. Phen.\,}{\bf {#1}},\,#2\,(#3)}
\def\epjc#1#2#3{{Eur. Phys. J.\,}{\bf C{#1}},\,#2\,(#3)}
\def\mpla#1#2#3{{Mod. Phys. Lett.\,}{\bf A{#1}},\,#2\,(#3)}
\def\etal{{\it et al.}}

\end{document}